\documentclass[12pt,preprint]{aastex}
\usepackage{xcolor}
\usepackage{ulem}

\newcommand{\median}{\mathop{\mathrm{median}}}
\newcommand{\abs}{\mathop{\mathrm{abs}}}

\begin{document}
\title{ROBOSPECT: Automated Equivalent Width Measurement}
\author{Christopher Z. Waters}
\affil{Institute for Astronomy, University of Hawaii at Manoa, 2680 Woodlawn Drive, Honolulu, HI 96822}
\email{watersc1@ifa.hawaii.edu}
\author{Julie K. Hollek}
\affil{University of Texas, Department of Astronomy, 2515 Speedway, Stop C1400, Austin, Texas 78712-1205 }
\email{julie@astro.as.utexas.edu}

\begin{abstract}
We present ROBOSPECT, a new code to automatically measure and deblend line
equivalent widths for both absorption and emission spectra.  We test
the accuracy of these measurements against simulated spectra as well
as manual equivalent width measurements of five stellar spectra over a
range of signal-to-noise values and a set of long slit emission
spectra.  We find that ROBOSPECT accurately matches both the synthetic and
manual measurements, with scatter consistent with the expectations
based on the data quality and the results of \cite{eqw_err}.
\end{abstract}


\section{Introduction}\label{intro}

Chemical abundance determination from stellar spectra is largely
achieved in one of two ways: synthetic spectra matching or
measurement of the equivalent widths of important spectral features.
The calculation of synthetic spectra is complicated due to the physics
that goes into the models, from the simplest set of atmospheric
parameters to the much more complicated considerations required for
effects such as non-local thermodynamic equilibrium.  Thus, it is
difficult to make direct comparisons between studies, not only due to
the different methods of dealing with the physics of stellar
atmospheres, but also the inclusion or exclusion of various features
(e.g., sources of broadening, non-local thermodynamic equilibrium
treatments) that can contribute to the formation and shape of a
spectral line.  

The alternative to this synthesis is to measure the equivalent width
of a set of lines.  The equivalent width is a geometric measure that
describes the total flux of a line as a rectangle with the same area
as the line.  By requiring the rectangle to have the height of the
normalized continuum, the width directly corresponds to the line flux.
The equivalent width can be measured without any prior knowledge of an
object or consideration for what input physics will be included in the
analysis.  However, once the equivalent width is measured, further
analysis with that information will then generally require the same
considerations for physics included in the models used for
spectral synthesis.

We have written a new equivalent width fitting program, ROBOSPECT, to provide
a consistent way to measure spectral line strengths.  By automating
these measurements, the data from current and future large-scale
studies can be quickly processed without concern about changes in fit
quality due to individual, manual measurements.  ROBOSPECT is designed to be
modular and open source, using freely available software and library
packages, to enable users to quickly adapt the code to their data
requirements.

Although ROBOSPECT was originally designed for the study of absorption lines
in stellar spectra, the algorithms are general enough to allow it to
be used to measure the equivalent widths of lines from emission
spectra.  Emission lines formed in rarefied gas can be used as
diagnostics for temperature, metallicity, density, and ionization
structure in astrophysical objects such as planetary nebulae
(e.g.~\citealt{pne}), active galactic nuclei (e.g.~\citealt{agn}, and
HII regions (e.g.~\citealt{h2r}).  In many instances, the useful value
is not the equivalent width, but rather the total line flux; however,
this is simply a mathematical conversion.

In this paper, we describe ROBOSPECT, an automated equivalent width
measurement program for absorption and emission lines.  In
Section~\ref{algorithms}, we outline the fitting algorithm that
determines the equivalent widths.  We detail the accuracy and
performance of ROBOSPECT on simulated and real data in
Section~\ref{accuracy}.  In Section~\ref{opp}, we discuss the features
of some other publicly available equivalent width measurement programs
and summarize our results in Section~\ref{conclusions}.

\section{Algorithms}\label{algorithms}

\begin{figure}
  \epsscale{0.5}
  \plotone{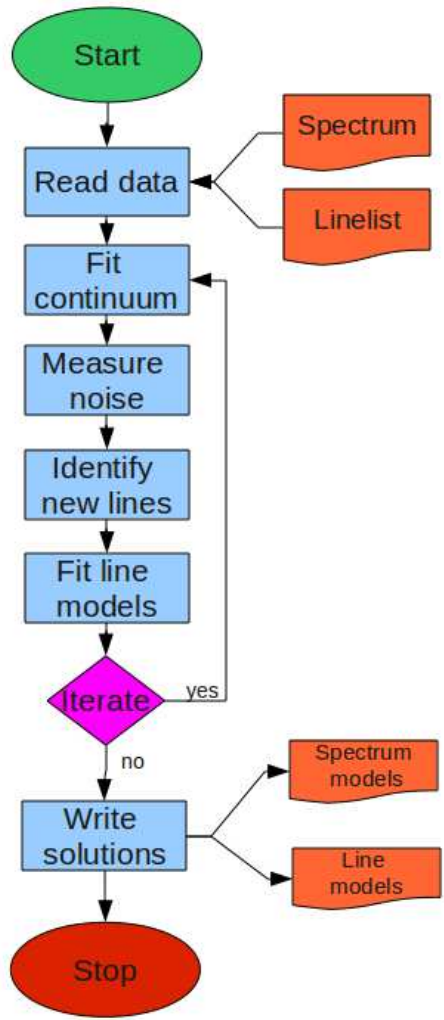}
  \caption{Diagram of ROBOSPECT steps used to fit the spectral components.}
  \label{fig:flowchartA}
\end{figure}

The goal of ROBOSPECT is to generate an accurate model fit to an input
spectrum, with a minimum of user interaction.  This helps ensure that
all fits are performed identically, regardless of the user or their
``personal fitting equation\footnote{The ``personal equation'' is a
  term that we use here to describe the phenomena where each
  spectroscopist fits lines in her own particular way in how she deals
  with continuum placement, fitting in the wings versus the core of
  the line and so forth.}''.  To that end, ROBOSPECT constructs automated
fits, with the addition of only a small number of user-defined
parameters that control the methods used in fitting.  In addition, ROBOSPECT
returns a set of flags for each fit performed, to allow the user to
note any quality issues, allowing poorly fit lines to be identified
largely from the resulting line catalog. Figure \ref{fig:flowchartA}
presents a diagram of the program structure to illustrate the order in
which spectrum components are measured.

The only required input to ROBOSPECT is a FITS or ASCII spectrum file,
listing the flux at each wavelength.  Uncertainties in the fluxes can
be supplied as well, but are not used in the fitting process, as they
tend to over-constrain the fits (see below).  Multi-order spectra are
supported, with each order fit independently. This prevents
discontinuities between the orders of a combined spectrum from
inducing errors in the continuum model.  This also allows orders that
overlap in wavelength space, such as those obtained with an echelle
spectrograph like the High Resolution Spectrograph on the Hobby-Eberly
Telescope, to be fit without merging the spectra.  A linelist of known
lines that should be fit can also be specified, providing the expected
line center.  These specified lines will always have a line fit
attempted, even if there is no obvious feature in the spectrum.  This
forced measurement method allows a consistent set of lines to be
measured across a large number of spectra, yielding upper limits if
the presence of an individual line is not obvious.

\subsection{Spectrum Models}
\label{sec:spectrum models}
\subsubsection{Continuum Normalized Spectra}

ROBOSPECT models spectra based on the assumption that all spectra are
comprised of three components: the continuum level, $C(\lambda)$, the
line solution relative to that continuum $L(\lambda)$, and an error
component $E(\lambda)$ that contains the deviation between the true
spectrum and the current model.  By iterating the fitting of these
components, we can ensure that the line and continuum solutions are not
biased by the other.  These components are combined in different ways
to measure line strengths, depending on the type of spectrum to be
fit.  For a continuum normalized spectrum (where the mean input
continuum is nearly unity), we assume that any low spatial frequency
deviations from unity are due to a poor normalization correction, and
high spatial frequency deviations are a combination of unknown and
unfit lines and the $E(\lambda)$ component.  In this case, we
construct a model of the final spectrum,$S(\lambda$), as:

\begin{equation}
  S_{continuum\ normalized}(\lambda) =
  C(\lambda) * \left(L(\lambda) + 1.0 + E(\lambda)\right)
\end{equation}

\begin{figure}
  \epsscale{1.0}
  \plotone{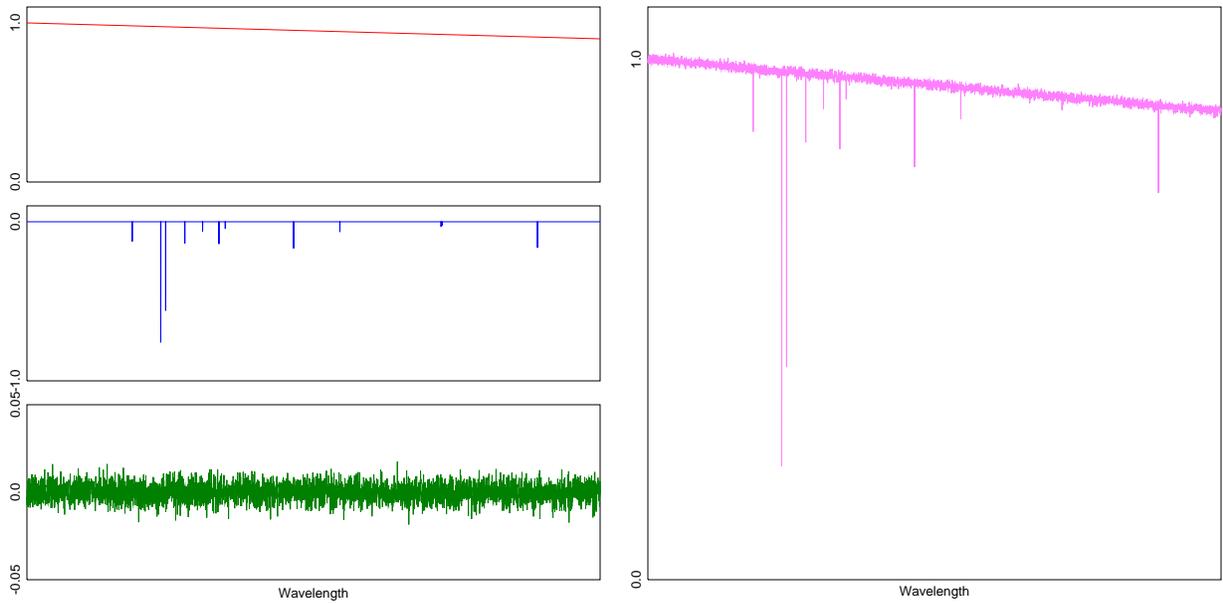}
  \caption{Simple diagram for the individual components that are used
    to make up the observed spectrum.  The left panels show the
    continuum (red), line (blue), and noise (green) models.  The right
    panel contains the expected observed spectrum, which is the
    product of the continuum model and the sum of the line and noise
    components.
    \label{fig:cartoonnormalized}}
\end{figure}

This relation ensures that in a spectrum with the continuum
normalization perfectly performed, the line model and noise add
directly to modulate the spectrum around unity.  The measured
continuum is assumed to represent imperfect
normalization, and so is multiplied to the expected final spectrum.
Figure \ref{fig:cartoonnormalized} demonstrates an example of how
these components combine to form the observed spectrum.

Using this spectrum model, we can solve Equation 1 for the continuum
component, by letting the input spectrum be the expected final
spectrum solution.  We can then remove the lines from the data, based
on previous iterations of the fitting routine.  For the initial
iteration, we have no knowledge of the correct line solution, and so
expect the continuum level to be slightly biased by strong lines.
However, as further iterations are performed, this bias decreases,
improving the continuum solution.  We ignore the error component when
measuring the continuum.  As the errors should cause deviations in
both directions from the true continuum, they should therefore not
bias the result.  By subtracting the current line model from the input
spectrum, we assume that we are simply left with the continuum
structure (modulated by the unknown error component).  For a
continuum normalized spectrum, we must first rescale the individual
line measurements to match the possibly imperfect input spectrum
continuum normalization:

\begin{eqnarray}
  C_{continuum\ normalized}(\lambda) &=&
  f_C\left(S_{input}(\lambda) - C(\lambda) * L(\lambda) \right) 
\end{eqnarray}

The function $f_C$ denotes some smoothing operation that returns the
continuum model, given the input of a spectrum with all lines
removed.  ROBOSPECT defaults to a simple
median boxcar filter for $f_C$, which can generally fit continuum
models that are smooth and do not have discontinuities.  Other continuum smoothing functions are supported, including a spline-based method
that can be used when the spectrum does have known discontinuities,
such as those due to the combination of echelle orders into a single
spectrum.  Functional models, such as a blackbody or power-law model, can also be
chosen for $f_C$, in which a known analytic function is fit to the
line-subtracted spectrum.

With the continuum modeled, we can start to investigate the effect
that the unknown $E(\lambda)$ has on the data.  This error is not, in
general, drawn from the per-pixel noise value.  As ROBOSPECT only attempts
fits for lines in the linelist, any unknown line will create a
deviation between the best model spectrum and the input data.  Because
of this, ROBOSPECT estimates a noise for each pixel that represents the
observed scatter in the spectrum, which includes the contribution from
small, unfit lines.  This measured noise is generally higher than the
expected Poissonian noise, due to this extra contribution.  This noise
is used both in the line fitting stage to determine how well the model
matches as well as in the internal line finding stage.  Similar to the
continuum solution above, we find a noise solution, $E(\lambda)$:
\begin{eqnarray}
  E_{continuum\ normalized}(\lambda) &=&
     f_E\left( \left(S_{input}(\lambda) / C(\lambda)\right) - L(\lambda) - 1.0\right) 
\end{eqnarray}

This solution removes the current best fitting continuum and line
models from the data, leaving only what should be considered the error
component.  As above in Equation 2, the function $f_E$ attempts to
model the underlying noise that this error component is drawn from.
This is also generally a filtering process, with the default filter
using the median absolute deviation statistic ($MAD = \median(\abs(x_i
- \median(x_i)))$).  This value is then converted to an equivalent
Gaussian $\sigma$ by noting that for a Gaussian distribution, the MAD
is half of the interquartile distance, and therefore this equivalent
$\sigma$ is equal to $1.4826 MAD$.  Using this statistic instead of a
directly measured standard deviation allows this estimate of the noise
to be robust against the influence of outliers, such as those that may
be left behind by large unmodeled spectral features.

At this point in the fitting process, ROBOSPECT can attempt to identify
potential lines that are not yet in its linelist.  This is done by
looking for local peaks that are more than a fixed user-specified
number of $\sigma$ deviant from the continuum subtracted level:

\begin{eqnarray}
  Z_{continuum\ normalized}(\lambda) &=&
     \left(\left(S_{input}(\lambda) / C(\lambda)\right) - 1.0\right) / E(\lambda) 
\end{eqnarray}

We expect that a large fraction of lines specified by the linelist
will be significant in this measurement.  Therefore, after finding all
the local peaks that are above the threshold, we merge this list of
peaks with those already in the linelist and remove the duplicates.
The current line model is not subtracted from the data before looking
for new lines, as the model residual can introduce significant local
peaks that are not necessarily real.

ROBOSPECT constructs a line solution model using the same methods as when it
finds lines, except without normalizing by the spectrum noise:
\begin{eqnarray}
  L_{continuum\ normalized}(\lambda) &=&
     f_L\left(\left(S_{input}(\lambda) / C(\lambda)\right) - 1.0\right) 
\end{eqnarray}     

Again, $f_L$ is a function that represents the fitting process
performed for each line in the current list (whether supplied by the
user or found automatically).  This fitting process can use various
methods, and can use different models for the line profile shape, as
described in Section \ref{sec:line models}.

\subsubsection{Flux Calibrated Spectra}

For flux calibrated spectra, the same three components are used to
describe the input spectrum.  However, these components are combined
in a different manner than is used for the continuum normalized
spectra.  For flux calibrated spectra, we assume that the continuum
solution combines with the line solution additively and not
multiplicatively as in the case of the continuum normalized spectra.
Therefore, the spectrum model for flux calibrated data is:

\begin{equation}
  S_{flux\ calibrated}(\lambda) =
  C(\lambda) + L(\lambda) + E(\lambda)
\end{equation}

From this model, we can derive relations for the various components as
was done above for the flux normalized spectra.
\begin{eqnarray}
  C_{flux\ calibrated}(\lambda) &=&
     f_C\left(S_{input}(\lambda) - L(\lambda)\right) \\
  E_{flux\ calibrated}(\lambda) &=&
     f_E\left( S_{input}(\lambda) - C(\lambda) - L(\lambda)\right) \\
  Z_{flux\ calibrated}(\lambda) &=&
     \left(S_{input}(\lambda) - C(\lambda)\right) / E(\lambda) \\
  L_{flux\ calibrated}(\lambda) &=&
     f_L\left(S_{input}(\lambda) - C(\lambda)\right) 
\end{eqnarray}

The same fitting and estimation algorithms can be used on both kinds
of spectra, as we isolate the component being measured in both
cases. Due to the large dynamic range possible for flux calibrated
spectra, it is useful to calculate the continuum and noise models
using the logarithm of the flux levels.  This prevents a large
continuum gradient from being dominated by the very largest values.

\subsection{Line fitting}\label{sec:line models}

\begin{figure}
  \epsscale{0.5}
  \plotone{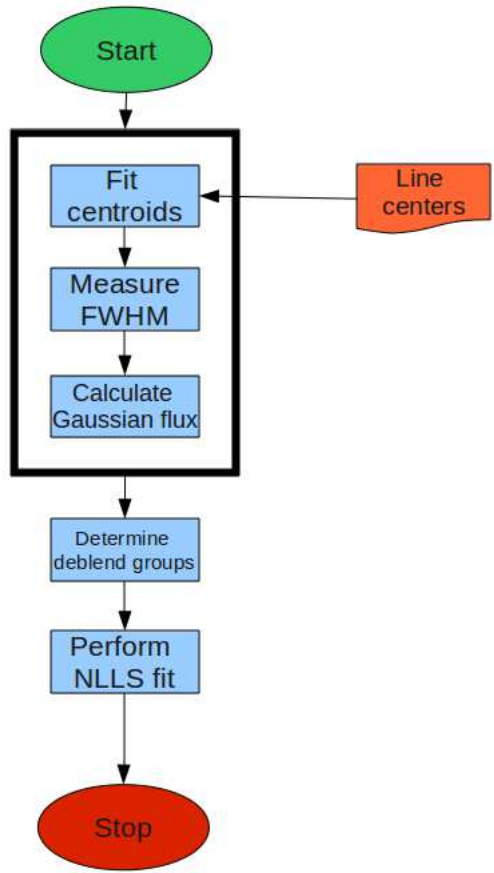}
  \caption{Diagram of the steps used by ROBOSPECT to fit individual lines.}
  \label{fig:flowchartB}
\end{figure}

Figure \ref{fig:flowchartB} shows a flow chart of the steps used in
fitting individual lines.  Before doing more complicated fitting, an
initial guess for the fit parameters is constructed for each line, under the
assumption that the line profile is a Gaussian.  The line center is
estimated from the centroid around the line peak.  The FWHM
is calculated by interpolating the line profile at the half-peak
value.  As heavily blended lines may have excessive contribution from
other lines at the half-peak value, we check this value with the width
calculated from the 75\%-max peak value.  In the case of large
differences in the expected Gaussian $\sigma$ between these two
values, we take the smaller value to minimize the possible effect of
blending.  Finally, total line flux, F, is estimated for a simple
Gaussian model from the peak flux and the FWHM ($F = P \sigma \sqrt{2
  \pi}$).  After this initial fit is performed for all lines, a set of
deblending groups is determined.  If the line centers are closer than
a given number of line widths (taken by default as six times the
Gaussian model $\sigma$), then they are assigned to the same
deblend group.  This deblend group is then simultaneously fit as a single
unit, to allow for the contributions of all the lines at a given
wavelength.

The full line fitting to find the best fitting set of model parameters
is done via non-linear least-squares minimization of $\chi^2=
\sum_{\lambda}\frac{L(\lambda) - \sum_{i} Q_i(\lambda)}{E(\lambda)}$
for each deblending group, where $Q_i(\lambda)$ is the component of
the $i$th line at wavelength $\lambda$.  Different line model choices
can made for $Q$, as described in Section \ref{subsec:line profile
  functions}.  Once convergence is reached for all the lines in one
deblend group, those values are recorded and the fitting moves to the next
deblend group, until all lines over the entire spectrum have been fit.

After fitting all lines, a set of quality checks are performed to
ensure that only valid fits are retained in the final model.  First,
any line that did not converge within a fixed number of fitting
iterations is flagged and rejected from further consideration.
Similarly, lines that have final fit values that are non-finite are
flagged and rejected.  Finally, the total line solution is calculated
by summing only the valid lines, further rejecting any lines that do
not reduce the total $\chi^2$ value.

\begin{figure}
  \epsscale{1.0}
  \plotone{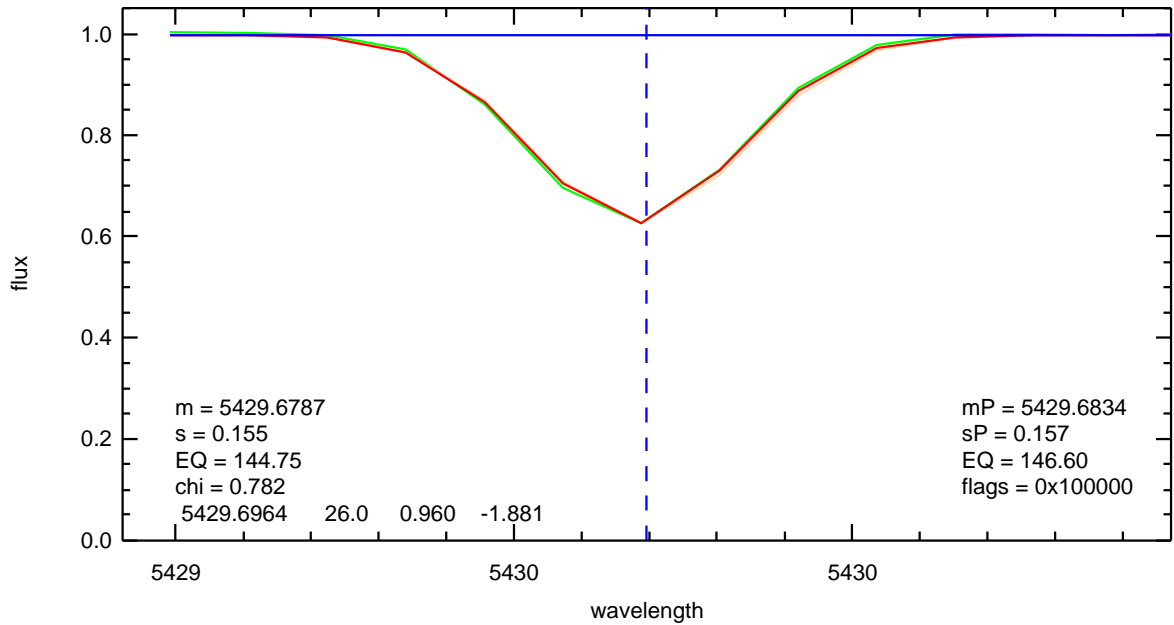}
  \caption{Example output fit from the S/N$\sim300$ star for Fe~I line
    $\lambda5429.68$.  The blue dashed line represents the line
    center, the blue solid line represents the continuum model, the
    green line represents the spectrum, and the red solid line represents the final
    line model. In the lower left corner, we list the final fit values
    for the line center, Gaussian $\sigma$, equivalent width, and fit
    $\chi^2$.  The lower right corner lists the measurements based
    solely on the preliminary fit calculated from the measured line
    FWHM, along with the hexadecimal value of the fit flags (in this
    case noting no issues with the line, only marking it as being fit
    with a Gaussian line model).}
  \label{roboplot}
\end{figure}

Once the line fitting has been finished and the line solution
accumulated, further iterations can be performed starting with these
models as an initial set of parameters.  As the continuum model is
based on the spectrum with lines removed, better line models result in
a smaller line residual, allowing for a better continuum model.  In
turn, with a more accurate continuum model, the line fits are more
reliable, with less noise as there are fewer residuals remaining.  In
Figure~\ref{roboplot}, we show an example output fit from the
S/N$\sim300$ star for Fe~I line $\lambda5429.68$.

\subsection{Line profile functions}\label{subsec:line profile functions}

The default assumption for all lines is that they have a profile shape
consistent with a simple Gaussian.  This is appropriate for most lines
used in a typical equivalent width-based stellar abundance analysis.
Lines that fall on the linear portion of the curve of growth have a
1:1 ratio between the line width and the absorbers, thus a small
change in the number of absorbers is easily reflected in the line
profile.  However, as the line leaves the linear portion of the curve
of growth, this ratio becomes smaller and the assumption of a
Gaussian profile begins to break down, resulting in wings that are
broader and contain a larger fraction of the total flux.  To account
for these strong lines, we allow the user to select other profile
functions that can better model these lines.

In the extreme case of the line shape generated solely by pressure
broadening, we support a Lorentzian model.  This is in general not
appropriate for most lines included in the typical analysis of a
stellar spectrum; however the Gaussian line shape is not appropriate
for certain lines from which abundances are derived (e.g., Mg b) and
for lines that are important in calculating the continuum (such as the
hydrogen lines).  Between the extreme Lorentzian and the Gaussian is the Voigt
profile, the result of the convolution of a Gaussian and Lorentzian
models.  This profile is somewhat difficult to directly calculate, and
so we utilize the Hjerting approximation given by
\citet{hjerting_reference} that describes the Voigt profile as a
polynomial approximation in $\eta$, the ratio of the Lorentzian width
to the Gaussian (Doppler) width.  This function reduces to the simple
Gaussian form when $\eta = 0$, allowing this function to be used on
all lines, with the degree of non-Gaussianity allowed to best match
the data.  This function is significantly more complicated than a
simple Gaussian, however, increasing the computation time required for
fitting.

\subsection{Wavelength solutions}

Internally, ROBOSPECT has no a priori knowledge of the quality of the
wavelength solution.  If a linelist is supplied, it is assumed that
the specified line centers are consistent with the peaks in the
spectrum.  If this is not true, then the lines specified may go unfit.
We have implemented some rudimentary ways to prevent this situation,
but these algorithms assume any errors in the wavelength solution are
small, and are not designed as a replacement for accurate wavelength
calibration.

Given that the linelist notes where expected lines should be found,
ROBOSPECT can use that information to evaluate the wavelength solution of
the supplied data.  By finding the local peak most consistent with the
input linelist, ROBOSPECT can check for systematic offsets between the
expected line center and the observed line center, and fit a
wavelength correction that makes these match.  This correction can
either be linear (the observed spectrum has an offset from
expectations by some small fixed $\delta\lambda$) or Doppler (the
observed spectrum has a velocity offset resulting in an offset
$\delta\lambda \propto \lambda$).  Once the correction has been
measured, this result is applied to shift the expected line centers to
match the data, not to change the wavelength solution of the data.
This choice is made to avoid issues interpolating the data onto a new
wavelength grid, and the possible resulting degradation in resolution.
The calculated correction is returned with the final fit, allowing the
user to use it to help re-reduce the data.

\section{Fitting Accuracy}\label{accuracy}

Although ROBOSPECT returns an error estimate for all measured parameters,
this is based on the fit $\chi^2$, and does not fully account for
the covariance between line parameters and continuum uncertainty.
Therefore, to confirm the fit quality and measurement accuracy, we
compare the results from ROBOSPECT to a sample of independently measured
lines.

\subsection{Simulated Spectra}
We performed our first test using a sample of 100 simulated spectra,
designed to be similar to a sample of high resolution,
continuum-normalized stellar absorption spectra.  Each spectrum covers
a range of 4000\AA{} centered at 6000\AA{}, with 0.1\AA{} spaced
pixels. A randomly chosen signal-to-noise ratio was selected
logarithmically between $S/N = 10$ and $S/N = 1000$ to represent the a
range of common data quality.  This S/N value was then used to add
random Gaussian noise around 1.0, representing the mean level of the
normalized spectra.  From this base, we then added 1000 equally spaced
Gaussian lines to the spectra.  The line equivalent widths were again
logarithmically sampled, from a range $1 < EQW < 600$ consistent with
what is observed in our spectra of real stars.  The FWHM values for
these simulated lines was drawn from the narrow range of $0.2 \AA <
FWHM < 0.4\AA$.  As the FWHM is largely a function of the resolution
of the data, and dependent on the instrument, there is little need to
probe a wide range of values.

In order to represent calibration errors that would shift the
continuum level, we multiplied the final spectrum with a spline model.
This spline model was defined by taking twelve equally spaced points
along the spectrum, and randomly selecting a value from the noise
expected by the signal-to-noise ratio.  These control points were then
interpolated using a cubic spline, and the simulated spectra scaled by
this curve.  This model assumes that the continuum level is more
robust in high signal-to-noise data, a reasonable expectation.

\begin{figure}
  \plottwo{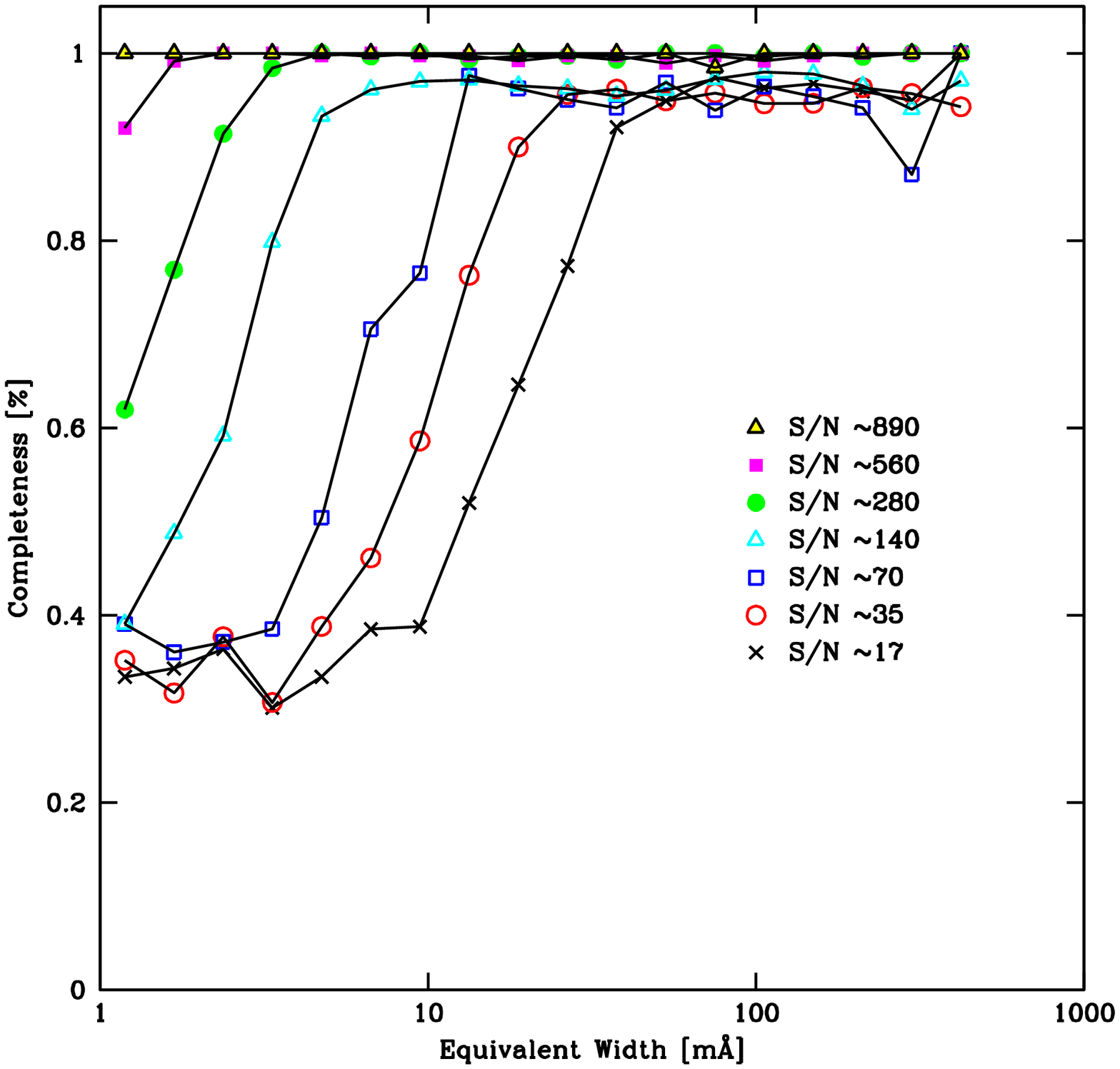}{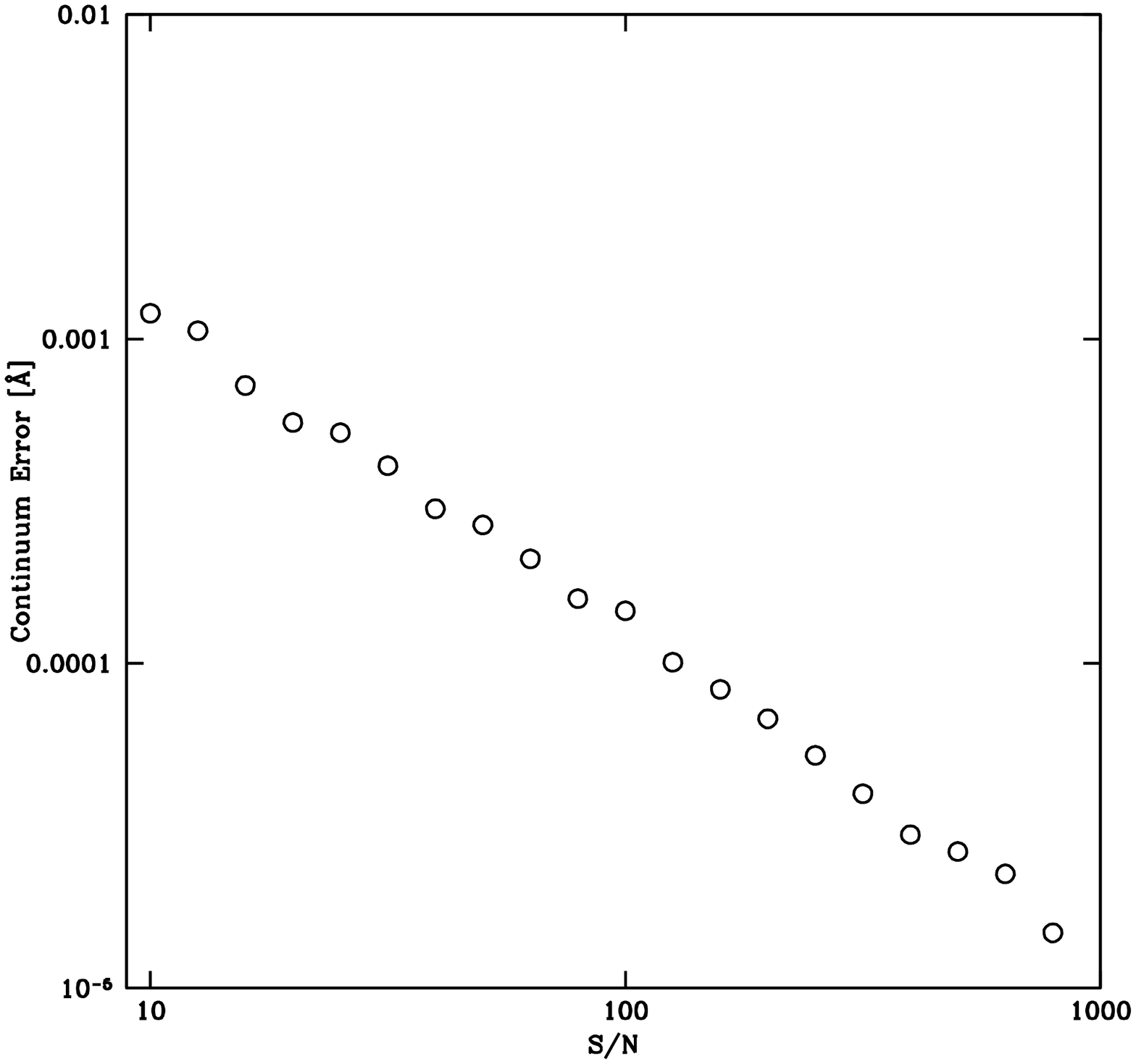}
  \caption{The left panel shows the completeness curves as a function
    of line equivalent width for the simulated spectra at a sample of
    signal to noise values.  The right panel shows the decrease in
    scatter between the measured and input continuum levels as the
    signal to noise of the spectrum increases. }

  \label{fig:spectrumquality}
\end{figure}

Once these simulated spectra were generated, they were processed using
ROBOSPECT, and the measured line parameters were compared with the known
simulation inputs.  Using the fit quality flags generated by ROBOSPECT, we
can exclude lines that were not fit correctly, and use this
information to determine the expected completeness of the fitting
process.  Figure \ref{fig:spectrumquality} shows the completeness as a
function of signal to noise and line equivalent width.  As is
expected, the completeness is worse for weak lines in low signal to
noise spectra.

One of the main issues with fitting spectral lines is ensuring that
the continuum level is estimated correctly.  As we know the input and
fit continuum levels, we can determine how much scatter to expect from
ROBOSPECT measurements.  This scatter is shown in
Figure~\ref{fig:spectrumquality}.  As expected, as the spectrum
signal-to-noise increases, the accuracy to which ROBOSPECT determines the
continuum also increases.  This continuum accuracy best represents a
lower limit to the uncertainty that is expected in real data.  Our
simulated spectra do not include the contribution that small faint
lines can have.  The presence of these lines in real data sets a noise
threshold in the continuum that exists in even the highest S/N spectra.

\begin{figure}
  \plotone{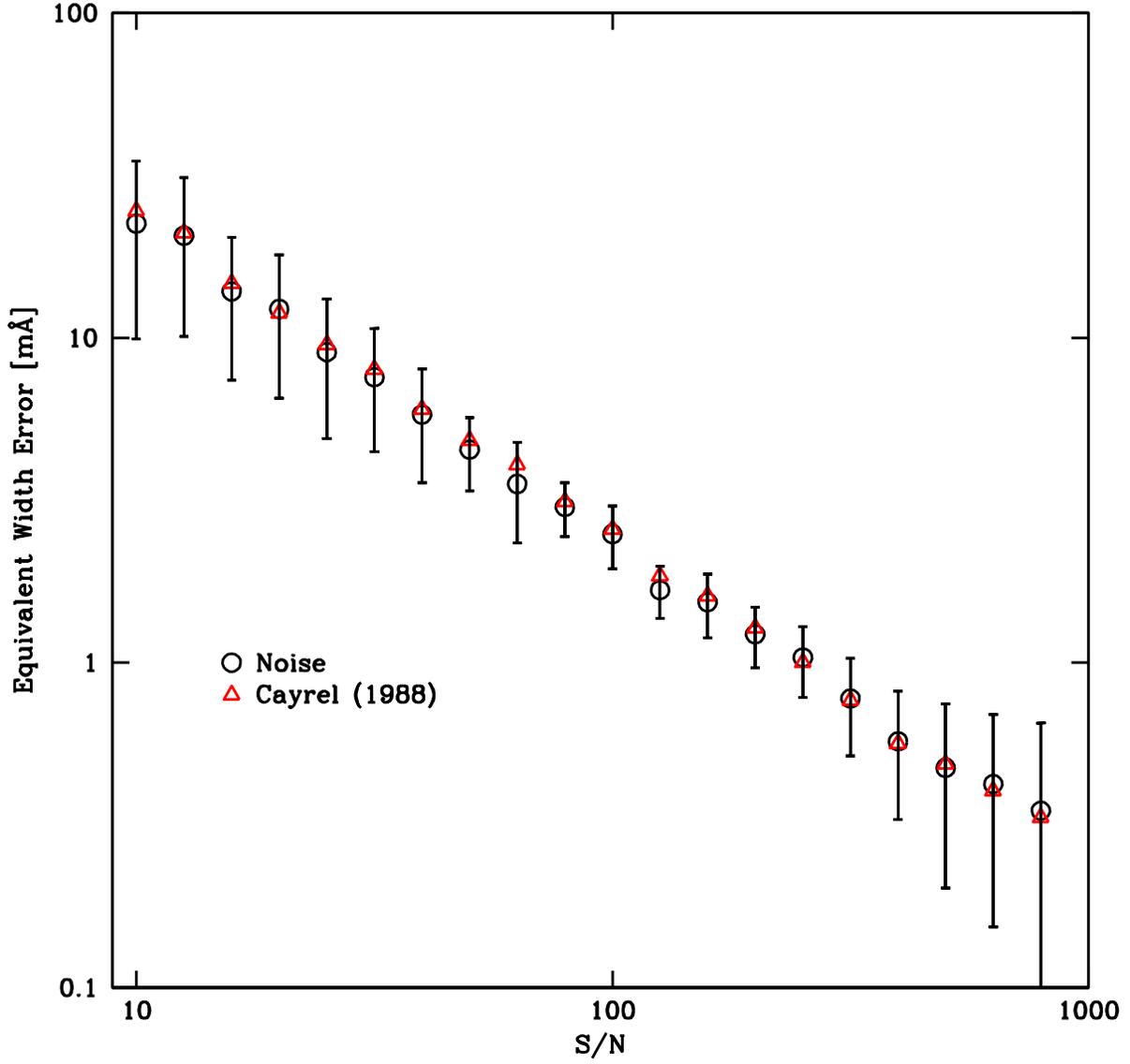}
  \caption{Observed equivalent width error distribution as a function
    of spectrum signal to noise.  For comparison, we also display the
    expected median scatter from \citet{eqw_err} using red triangles.
  }
  \label{fig:eqwquality}
\end{figure}

The goal of ROBOSPECT is to accurately measure the equivalent widths of the
lines in the spectrum.  Therefore, understanding the errors on these
measurements is essential.  From \citet{eqw_err}, we expect that the
scatter is a function of the line FWHM and the spectrum signal to
noise.  The simulated spectra provide a very large sample of 65,000
lines with valid line measurements to estimate the observed scatter in
the ROBOSPECT fits.  These results are displayed in Figure
\ref{fig:eqwquality}, along with the \citet{eqw_err} predictions.  As
mentioned before, the presence of faint lines in real data is likely to
increase the observed fit scatter, but based on this sample, it is
clear that the fits produced by ROBOSPECT are consistent with the expected
theoretical accuracy.

\subsection{CASH Star Comparison}

To test ROBOSPECT on a set of real stellar spectra, we compared the
equivalent width measurements made for 5 stars of S/N $\sim300$,
$100$, $65$, $45$, and $30$, listed in Table~\ref{teststars}.  These
stars were observed as part of the Chemical Abundances of Stars in the
Halo (CASH) project \citet{cash2}.  The CASH project is a study of 500
stars, observed at McDonald Observatory using the High Resolution
Spectrograph \citep{hrs} on the Hobby-Eberly Telescope \citep{het}.
The goal of this study is to analyze a statistical sample of stellar
spectra to determine chemical abundances and abundance trends in a
self-consistent manner. We are using ROBOSPECT to facilitate the analysis
for this large spectroscopic study.

These S/N values were chosen to present an ideal case, a good case,
the median S/N of the CASH study, a bad, and a very bad case.
Observational details of the CASH project are given by \citet{cash2}.
As these stars are part of a study of the stellar halo, they are of
low metallicity, ranging from $-3.0 <$[Fe/H]$<-2.4$.

These stars were observed at R$\sim15,000$ and have been measured
using a semi-automated IDL routine \citep{ew} which was also used for the stars
of equal resolution by \citet{cash2}.  Here we examine two of the most
common parameter settings used in ROBOSPECT: standard Gaussian line profiles
and Hjerting line profiles.  In both cases, we use the set of default
parameters listed in Table~\ref{defaultparam} to test the code, which
was run for five iterations of the main fitting loop.  The only
deviation from the default parameters is for the fits using the
Hjerting line profile shape.

Tables~\ref{gaussanal} and \ref{hjertinganal} show the results of
statistical tests that quantify the difference between the manually
measured equivalent width and the ROBOSPECT measured equivalent width.  We
also include the S/N, the iteration of fitting loop presented, as well
as the number of lines that had fits considered valid by ROBOSPECT.  To
prevent a small number of outliers from skewing the results, we also
list the median and the robust estimate of the standard deviation,
$\sigma_Q$.

\subsubsection{Gaussian Line Profile Statistical Analysis}
In the standard case, using the default parameters with a Gaussian
line fit, we find that ROBOSPECT matches the manual measurements, with
differences between the manually measured result consistent with the
expected \citet{eqw_err} estimate.  We show results for the test
spectra for five iterations in Table~\ref{gaussanal}.  We determined a
nominal error in the manually measured equivalent widths as calculated
from \citet{eqw_err} by basing the wavelength in the region near the
Mg b lines.  This region was chosen as it was the same one from which
the S/N was calculated.  We find the errors in equivalent width to be
1.00, 3.00, 4.61, 6.66, and 9.99 m\AA{} for the 300, 100, 65, 45, and
30 S/N cases respectively.  As with the simulated spectra, as the S/N
drops the accuracy with which ROBOSPECT matches the manual equivalent width
also drops.  In all cases but the 300 S/N case, the standard deviation
falls below this value.  Furthermore, the $\sigma_Q$ value is well
below the error in the manually measured equivalent widths, again with
the exception of the S/N 300 case. The likely cause for this exception
is the differences in continuum level chosen by the manual and ROBOSPECT
measurements.  The continuum level from ROBOSPECT passes through the median
of the spectra data, which tends to be slightly lower than the manual
placement, resulting in a small bias.

Increasing the number of iterations generally decreases both the
fitting bias (as measured by the mean and median) as well as the fit
scatter (based on $\sigma$ and $\sigma_Q$).  Usually, three iterations
are sufficient, with a minimal improvement in the scatter beyond this
limit of $\sim0.02$m\AA{}.  This improvement is small relative to the
expected uncertainty in the manual measurements, suggesting that
further iterations are not likely to improve the fit results.

\begin{figure}
  \plottwo{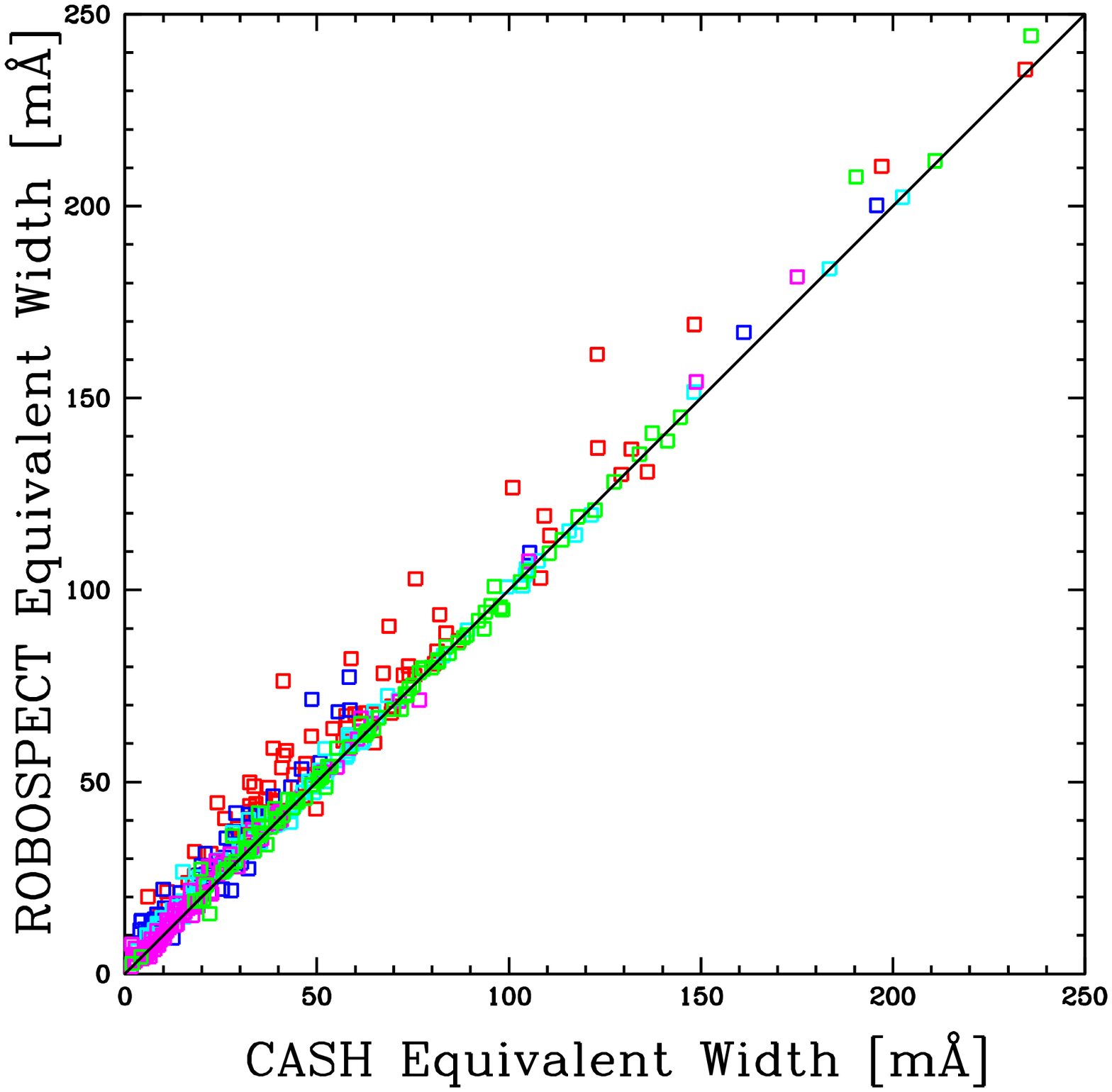}{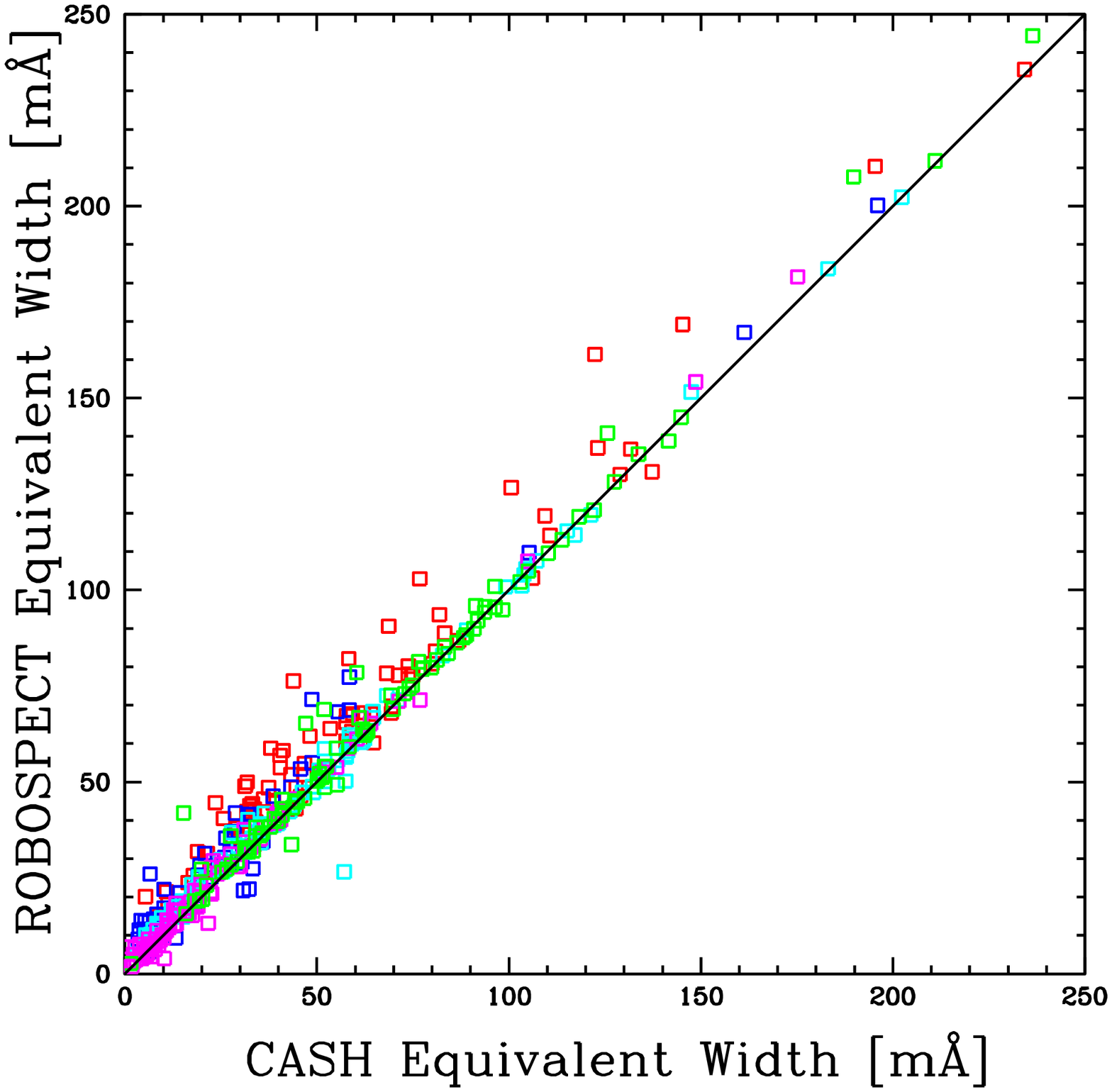}
  \caption{Fit comparison for Gaussian (left) and Hjerting (right)
    models at all S/N levels.  The green boxes are for S/N 300, magenta for
    S/N 100, cyan for 65, blue for 45, and red for 30.  The thick black
    line is the 1:1 ratio.
    \label{fig:compA}}
\end{figure}

The left panel of Figure~\ref{fig:allres} shows plots of the
residuals between the manually measured equivalent widths from
\citet{cash2} and the ROBOSPECT equivalent widths, with each S/N value
offset by 50 m\AA{}.  The left panel of Figure~\ref{fig:compA} shows
the equivalent widths from the manually measured equivalent widths
against those of ROBOSPECT for all S/N values.

Although the scatter between the measurements matches the
expectations, there is a clear S/N dependence in the bias between the
ROBOSPECT and manual measurements.  The most plausible reason for this is a
difference in the way that the continuum level is set between these
two methods. ROBOSPECT assumes that the true continuum level is ``hidden''
within the scatter in the spectrum data, and places the continuum
level to be consistent with the median value of this scatter.  In
contrast, a manual measurement is likely to set the local continuum at
the peaks to either side of the line, substantially higher than the
ROBOSPECT value (on the order of $\sigma_{continuum}$, the scatter in the
continuum level).  Because of this, the increasing deviation between
the manual measurements and the ROBOSPECT measurements as the signal to
noise decreases is not surprising, and represents the covariance
between continuum level and line flux.

\subsubsection{Hjerting Line Profile Statistical Analysis}
As some very strong lines ($\gtrsim$$-4.5$ in reduced equivalent
width\footnote[1]{Where reduced equivalent width is
  $\log_{10}$(Equivalent Width/Wavelength)}) may have a significantly
non-Gaussian line profile shape, we repeated the analysis of these
stars with the Hjerting line model instead of a Gaussian.  The
comparison with the manual equivalent width measurements is detailed
in Table~\ref{hjertinganal} over all five iterations.

\begin{figure}
  \plottwo{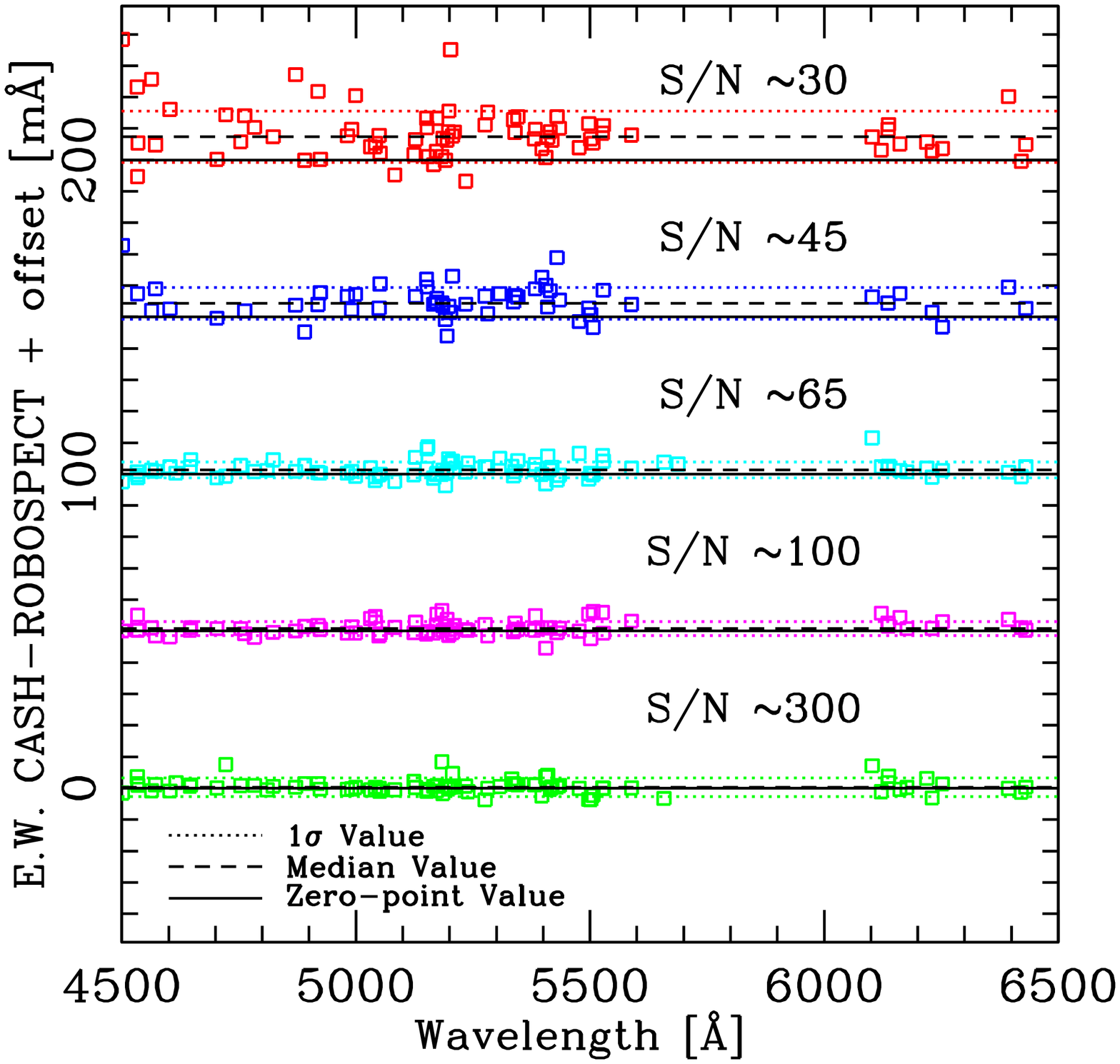}{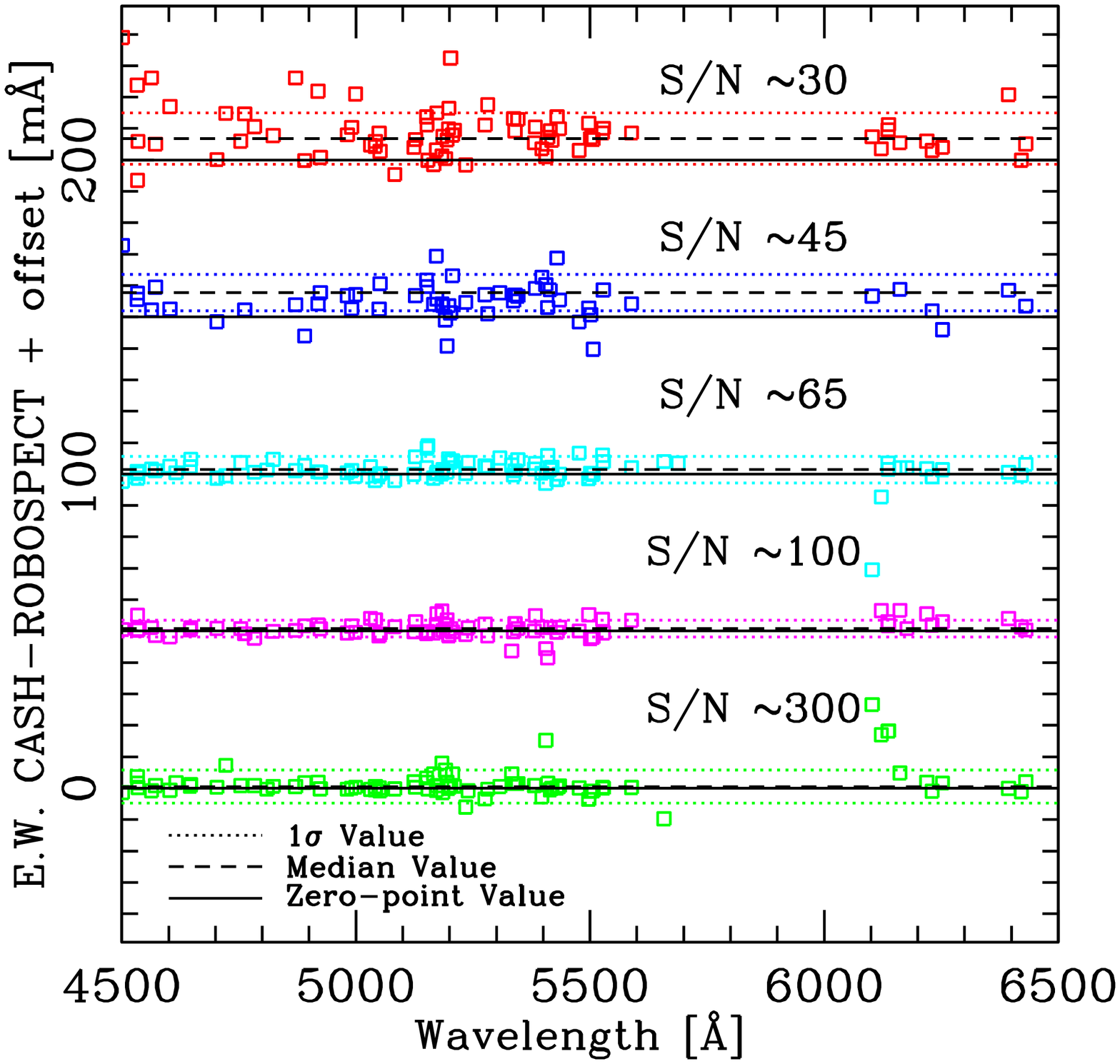}
  \caption{Plots of the residuals between the equivalent widths
    measured by the semi-automated IDL code and ROBOSPECT for all
    data, with 50 mA offsets for each S/N value.  On the left are the
    residuals for the Gaussian profile and on the right the Hjerting
    (Voigt) profile.  In each plot, the green points represent the S/N
    $\sim300$ data, the magenta points S/N$\sim100$ data, the cyan
    points S/N$\sim65$ data, the blue points S/N$\sim45$ data, and the
    red points S/N$\sim100$ data.  For each set of points, there is a
    color-matched dotted line which represents the relative 1$\sigma$
    value centered around the median value, denoted by the black
    dashed line.  The black solid line represents the zero-point for
    each S/N value.
    \label{fig:allres}}
\end{figure}

The right panel of Figure~\ref{fig:allres} shows plots of the
residuals between the manually measured equivalent widths from
\citet{cash2} and the ROBOSPECT equivalent widths, with each S/N value
offset by 50 m\AA{}.  The right panel of Figure~\ref{fig:compA} shows
the plot of the equivalent widths from the manually measured
equivalent widths against those of ROBOSPECT for all S/N values.  Upon
inspection of the statistical results, we find that the mean and
$\sigma$ values are significantly larger than those associated with
the standard Gaussian fits; however, when we evaluate the median and
$\sigma_Q$ values, we find that they are reasonably consistent.  As
the median and $\sigma_Q$ values are less sensitive to outliers, this
suggests that the added parameters of the Hjerting model may make the
code more susceptible to erroneous fits.  It should also be noted that
night sky lines can pollute the stellar spectrum around 6000\AA{},
which is where many of these outlier fits exist, as seen in the right
panel of Figure~\ref{fig:allres}.

\begin{figure}
  \plotone{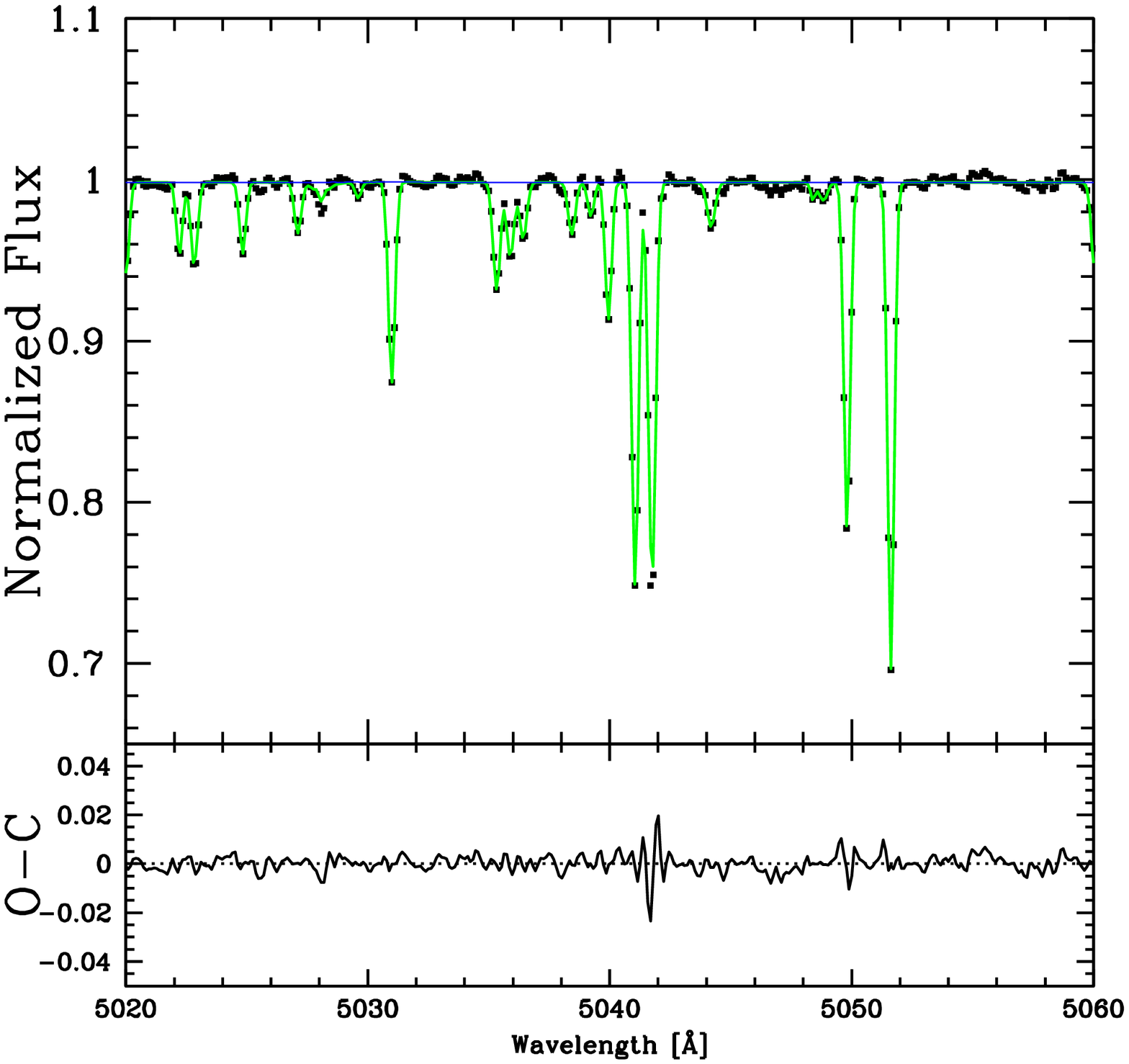}
  \caption{Plot of the line model (solid green line) and continuum
    model (solid blue line) against the input spectrum (black
    squares), with the residuals between the input spectrum and line
    model plotted at the bottom (solid black line) for the S/N 300
    case.
  \label{fig:300ocplot}}
\end{figure}

\begin{figure}
  \plotone{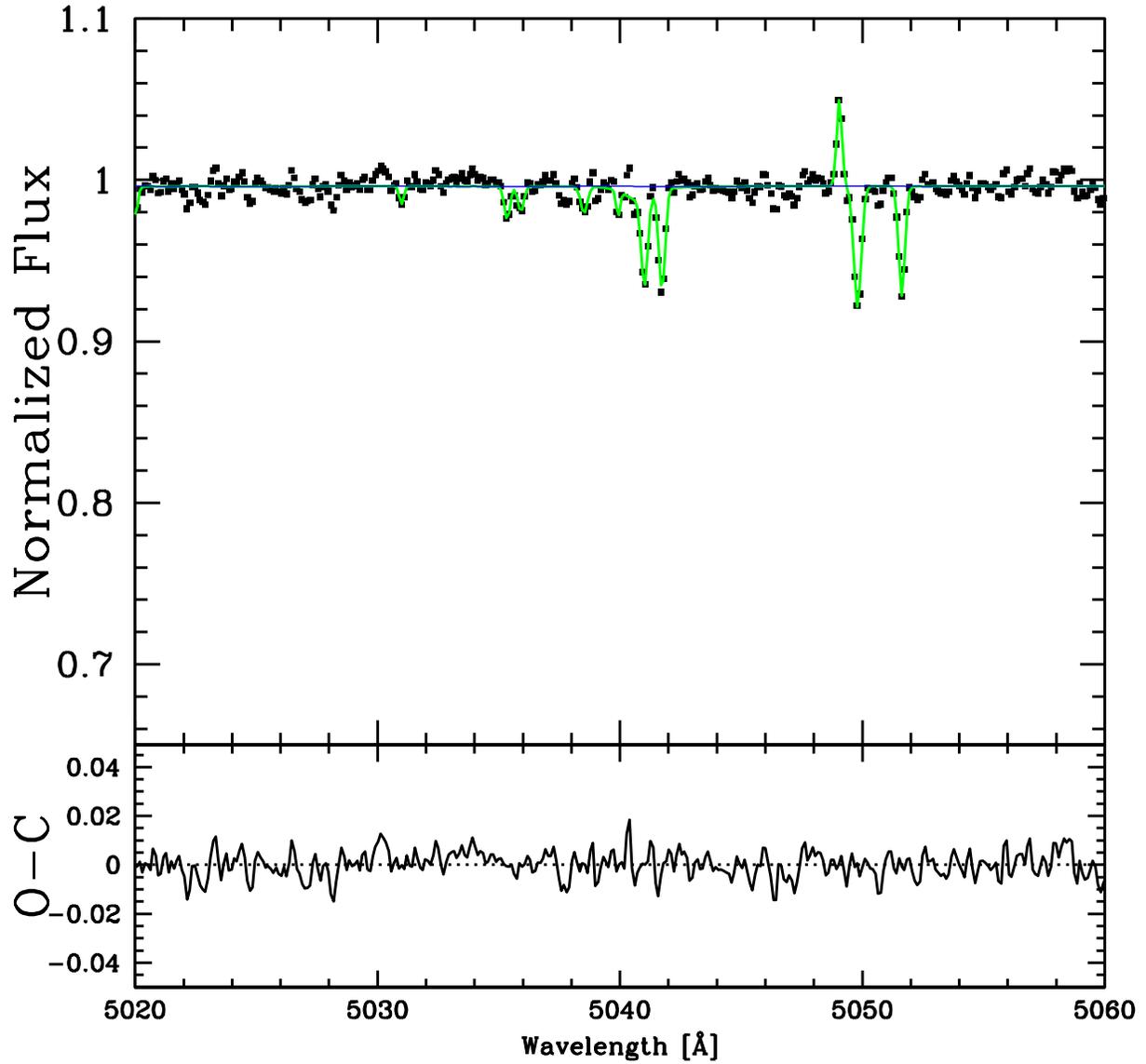}
  \caption{Plot of the line model (solid green line) and continuum
    model (solid blue line) against the input spectrum (black
    squares), with the residuals between the input spectrum and line
    model plotted at the bottom (solid black line) for the S/N 100
    case.  One noisy peak is fit as emission lines at $\lambda$5048.
  \label{fig:100ocplot}}
\end{figure}

\begin{figure}
  \plotone{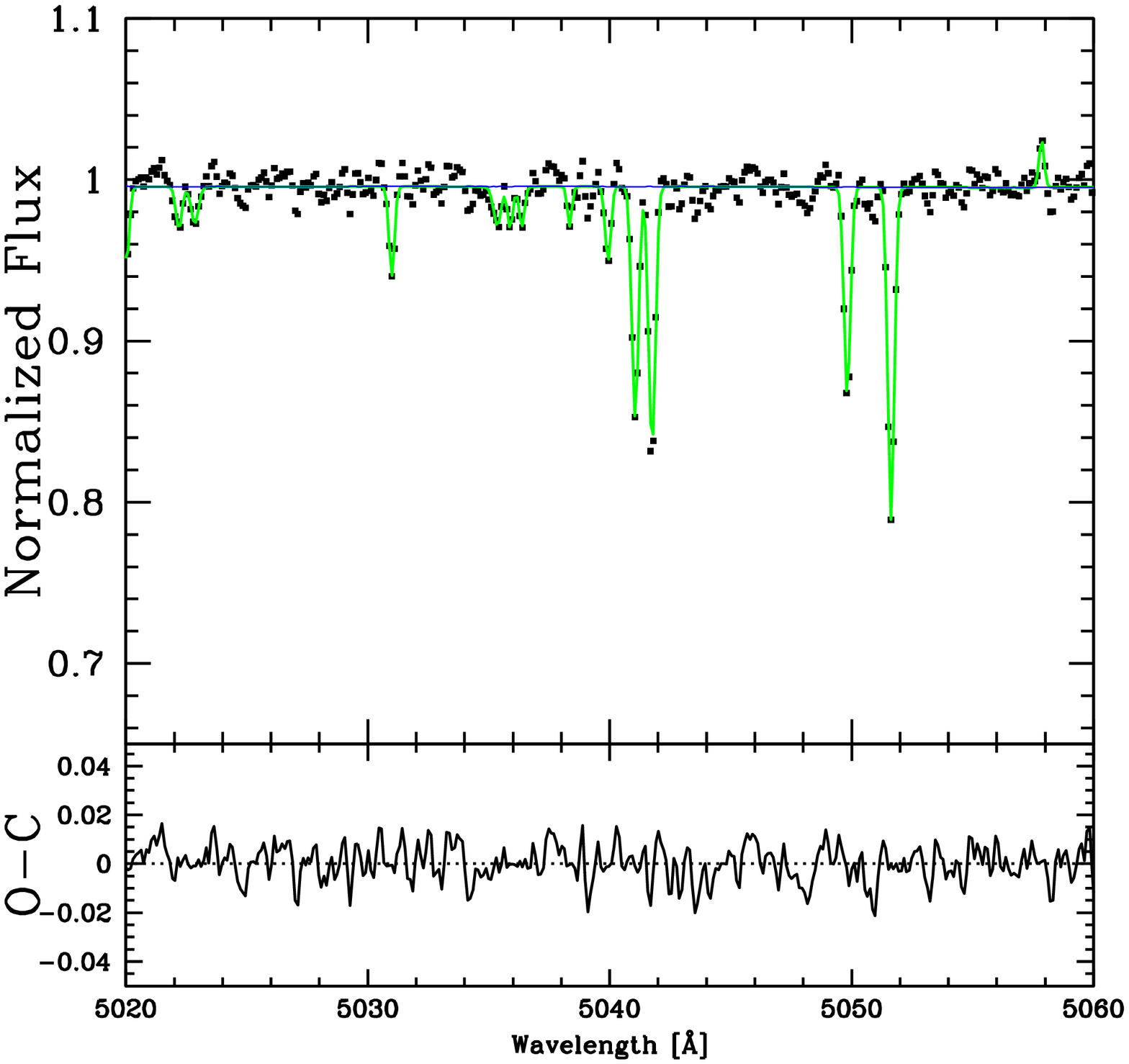}
  \caption{Plot of the line model (solid green line) and continuum
    model (solid blue line) against the input spectrum (black
    squares), with the residuals between the input spectrum and line
    model plotted at the bottom (solid black line) for the S/N 65
    case.
  \label{fig:65ocplot}}
\end{figure}

\begin{figure}
  \plotone{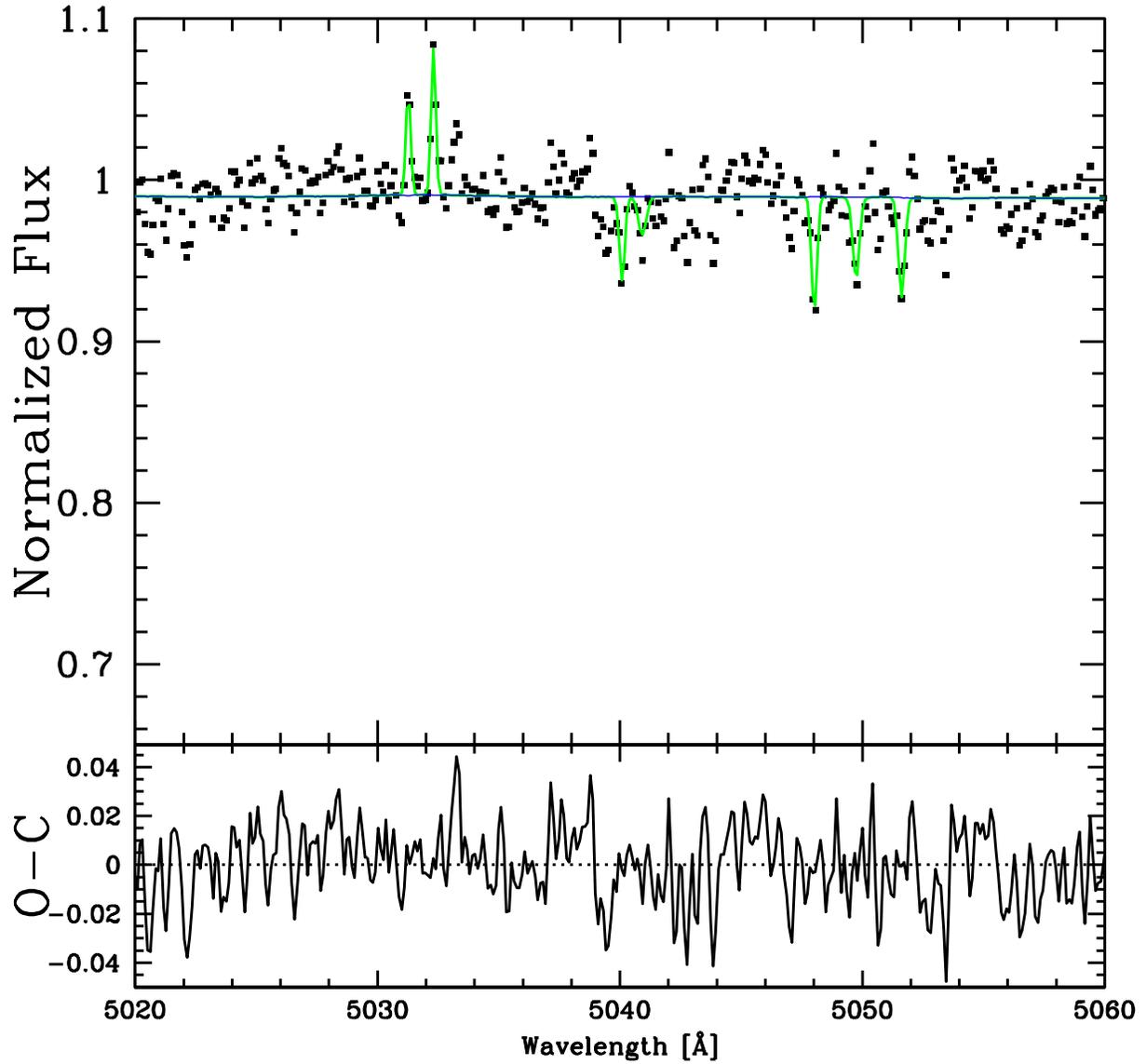}
  \caption{Plot of the line model (solid green line) and continuum
    model (solid blue line) against the input spectrum (black
    squares), with the residuals between the input spectrum and line
    model plotted at the bottom (solid black line) for the S/N 45
    case.  Two noisy peaks are fit as emission lines at $\lambda$5031
    and$\lambda$5032.
  \label{fig:45ocplot}}
\end{figure}

\begin{figure}
  \plotone{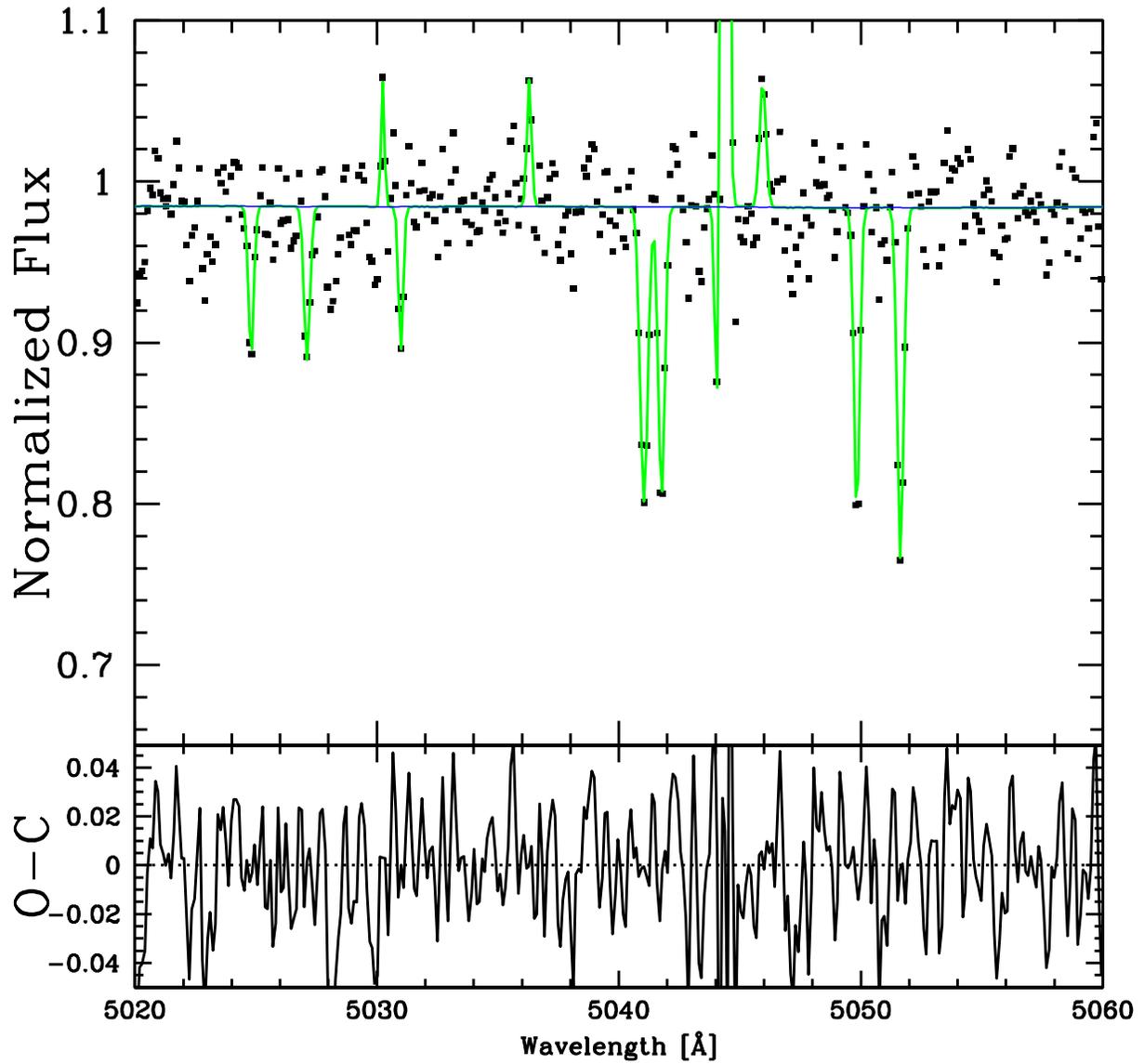}
  \caption{Plot of the line model (solid green line) and continuum
    model (solid blue line) against the input spectrum (black
    squares), with the residuals between the input spectrum and line
    model plotted at the bottom (solid black line) for the S/N 30
    case.  Three noisy peaks are fit at $\lambda$5030, $\lambda$5038,
    and $\lambda$5046 as emission lines.  A cosmic ray is also fit as
    an emission line at $\lambda$5044.  
  \label{fig:30ocplot}}
\end{figure}

Figures~\ref{fig:300ocplot}-\ref{fig:30ocplot} include a visualization
of the line model and model continuum plotted against the input
spectrum along with the residuals between the spectrum and line model
for each S/N value of the CASH stars presented for the Hjerting line
profile fits.  The wavelength region shown spans from 5520\AA{} to
5560\AA{}, which is also the width of the deblending region.  This
region includes a several lines which require deblending, along with
some individual features as well.  As the metallicity of each of these
stars is different, the line strengths vary as well, as strongly
evidenced by the $\lambda$ 5041.07 and $\lambda$ 5041.76 lines, two
prominent Fe features.  These example fits also show that strongly
deviant points, such as those caused by cosmic rays, are detected and
fit as emission features. These fits help minimize the influence that
these discrepant points have on the continuum model.

We also tested the Hjerting line profiles against the published
equivalent width measurement values for HD~122563, the S/N$\sim300$
star, from \citet{cayrel2004} and for HD~115444, the S/N$\sim65$ star,
from \citet{fulbright00}.  We find good agreement with both sets of
equivalent widths 0.953$\pm$5.796 m\AA{} and 1.768$\pm$2.621 m\AA{}
for HD~122563 and HD~115444, respectively.

\subsubsection{Statistical Analysis for a Pared Linelist for Both Gaussian and Hjerting Profiles}

In our tests, we used a linelist that was designed for metal-poor
stars. However, not all stars have every line in the list, and
artifacts such as cosmic rays and noisy spectral regions often make
certain lines that have been measured by ROBOSPECT undesirable for further
analysis.  Even when using manual programs to measure equivalent
widths, there are lines that simply cannot be measured well enough for
inclusion subsequent analysis such as stellar parameter determination.
The previous set of statistics included all lines that were measured
by ROBOSPECT and not flagged by the code as invalid.  As such, they may
include lines with these undesirable features.  We performed a
statistical analysis on a subset of these lines which have been
deemed suitable for stellar abundance analysis and present the results
in Table~\ref{cookedanal}.  We include only the statistical results
from the fifth iteration, as these are the final fit values.

In general, we find the same trend in these improved statistics: high
S/N values correspond with smaller mean differences, medians,
$\sigma$, and $\sigma_Q$ values.  By removing spurious lines, we find
better agreement with the manual equivalent width measurements.  Of
particular importance is that once the outliers are removed, the
Hjerting line profile fits match slightly better than the Gaussian
fits as compared to the manual measurements.  When accepting all lines
that ROBOSPECT deemed valid, the Gaussian fits performed better; hence, the
Hjerting profile is more susceptible to outliers, but should be
used when the user can remove particularly bad fits. Below we
highlight some statistics of note.

In the 45 S/N Gaussian case, the $\sigma$, mean, and median improve,
but the $\sigma_Q$ value increases.  This is likely because the line
fits that were deemed bad were less discrepant from the manual
measurements than some of the lines that were left in the re-analysis
of the statistics.  These seemingly ``very bad'' lines were left in
because the discrepancy in the measured equivalent widths is a result
of the different choices in the continuum level.  It is likely that
this creates a problem for the 45 S/N case and not the others
because the S/N is low enough to begin to hide weak lines in the noise
which would pull down the continuum level to a significant level, but
not so low that a large fraction of the lines are lost in the noise.

In the 65 S/N Hjerting case, we find that the mean increases in the
re-run after removing the bad fits.  For this run, the extreme
outliers were on the lower end, bringing the mean down closer to 0.
By removing the highly negative outliers, the mean increased; however,
the median, which is more robust against outliers decreased along with
the $\sigma_Q$ and $\sigma$ values as is expected.
                                                                            
\subsection{Emission Spectrum Comparison}

As discussed in Section~\ref{sec:spectrum models}, the method used for
fitting flux calibrated data is slightly different than that used for
normalized spectra.  To test the accuracy of this fitting mode, we
examined a set of long slit emission spectra of the Orion nebula.
These data were taken on the 1.5-meter Cassegrain Spectrograph at
CTIO.  Three gratings were used, providing a blue, red, and high
resolution spectrum for each pointing.  The characteristics of these
gratings is provided in Table \ref{tab:orion1.5m}.

The spectra were flux calibrated, and apertures defined to bin the
spectra to ensure an average signal-to-noise of $S/N \approx 10$.
This results in thirty binned spectra for each of the 24 input
images. A manual measurement of the lines was performed using the IRAF
task SPLOT, with the continuum and center set manually.  

\begin{figure}
  \plotone{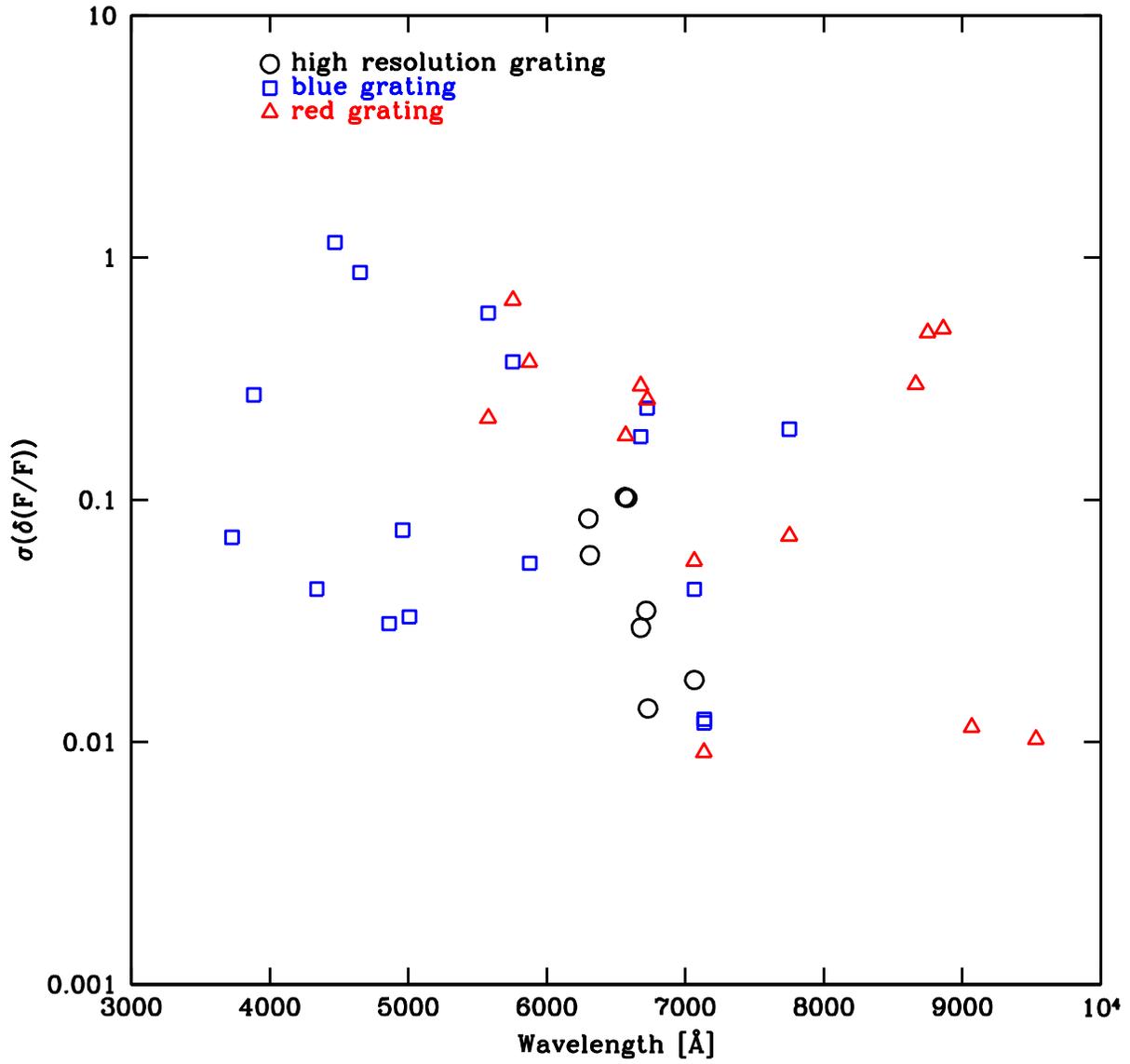}
  \caption{Observed relative flux scatter between the ROBOSPECT and manual
    SPLOT measurements for individual lines in the Orion emission
    spectra.  The three gratings are plotted separately, illustrating
    that the two methods are more consistent for the higher resolution
    data, which has fewer issues with blending.}
  \label{fig:orion_lines}
\end{figure}

ROBOSPECT was then used to measure each spectrum as well, using a single
list of lines and a Gaussian line model.  The relative flux difference
($(F_{ROBOSPECT} - F_{SPLOT}) / F_{ROBOSPECT}$) and the scatter are listed in
Table~\ref{tab:orion1.5m}.  For the majority of lines, the results
agree very well, again illustrating that ROBOSPECT produces measurements
consistent with other methods of spectral line measurement.  Figure
\ref{fig:orion_lines} shows the scatter observed in individual lines
for each grating.  It is clear from this that the scatter in the high
resolution spectra is much lower than for the other two gratings.  For
the lines with the largest difference, a comparison of the SPLOT and
ROBOSPECT measurements shows that ROBOSPECT fluxes are generally lower than those
measured by SPLOT.  This decreased flux is a result of ROBOSPECT deblending
neighboring lines that were fit with a single Gaussian in the SPLOT
analysis.  Therefore, this increased scatter is largely a result of
the manual measurement incorrectly modeling the data.

\subsection{Solar Spectrum Analysis}

The assumption that all lines can be fit by Gaussians or the Hjerting
model is certainly not valid for the strongest lines.  For these very
broad lines, ROBOSPECT will underestimate the line flux as the true line
shape deviates from the analytic forms.  Such lines are not commonly
used in the calculation of stellar properties in studies like the CASH
project, but knowing the point at which the analytic models break down
is essential.

This break down point can be calculated with a well-calibrated real
spectrum that contains a large number of lines that vary over many
orders of magnitude in equivalent width. Furthermore, using a spectrum
with a very high signal-to-noise ratio and resolution reduces the
possibility that faint lines blended into the wings of strong lines
will make those strong lines deviate from the analytic form.  The
solar spectrum presented by \citet{wallace_solar} satisfies all of
these requirements, and in addition, has a well-calibrated continuum
level that we can take as fixed to reduce the influence that the ROBOSPECT
continuum level can have on the line measurements.

From this high quality spectrum, we selected the region between 3825
and 4561 \AA{} to study.  This region contains many lines, with a
range of equivalent width values from 1 to 1000 m\AA{}.  To do the
comparison, we first ran ROBOSPECT with the Hjerting line model selected.
For each of the well-fit lines in the output catalog, we attempted to
construct an independent non-parametric measurement.  As these lines
may still have contaminating blended lines, we constrain the
non-parametric model curve in two ways.  First, the line must be
symmetric about the central wavelength.  This constraint helps remove
these blends, as we can use symmetry to exclude the influence on the
line under consideration.  Secondly, we demand that a line
monotonically approaches the continuum.  This also helps prevent
issues with blending, and ensures that the line has a finite total
flux.  Once each line shape was defined by this non-parametric model,
we used a direct integration of the line flux to obtain the total
equivalent width.

Unsurprisingly, this non-parametric model generally produces larger
equivalent widths than the standard Hjerting fits, as it can probe
further out into the wings of the lines.  For weak lines, this effect
is minimal, as any wings are small and reasonably fit by the analytic
model.  For lines with non-parametric equivalent widths less than $EQW
\sim 100$, the residuals between the non-parametric model and Hjerting
model are largely random, with an average deviation of $6.1\pm12.7
m\AA$.  Above this limit, however, the deviations and scatter increase
dramatically, as the lines become increasingly dominated by the wings
instead of the core.  This change is poorly modeled by the Hjerting
function, resulting in a larger average deviation of $52.6\pm51.1
m\AA$.  Separating out the measurements above $EQW = 100$, we can fit
a line to the deviations and note that the error in the Hjerting fits
is largely linear with the true equivalent width, such that $\Delta
EQW = 40.22 - 0.56 EQW$.

\section{Other Equivalent Width Measurement Software}\label{opp}

The future of astronomical study is in large collaborations,
specifically large surveys, such as those like the Sloan Digital Sky
Survey (SDSS) \citet{sdss} and the subsequent Sloan Extension for
Galactic Understanding and Exploration (SEGUE) \citet{segue}, both of
which included a large spectroscopic component.  For these surveys,
and for any statistical stellar sample, it is necessary to analyze the
spectra in a self-consistent manner so that the resultant abundance
trends and frequencies are meaningful.  SDSS and SEGUE employ the
Sloan Stellar Parameter Pipeline (SSPP) \citep{sspp} for such analysis
of its medium resolution spectra.  However, for studies with higher
resolution data, there exist several codes to measure equivalent
widths, although are truly fully-automated.  ROBOSPECT has been
written to fill this niche, and perform automated fits to high
resolution spectra.

A variety of very good publicly available programs exist, and can
largely be divided into three categories: manual, semi-automated, and
fully automated methods.  Manual programs, such as SPLOT in
IRAF\footnote{IRAF is distributed by the National Optical Astronomy
  Observatories, which is operated by the Association of Universities
  for Research in Astronomy, Inc., under cooperative agreement with
  the National Science Foundation.}, require continuous user
interaction, and due to this, the measurements can be significantly
influenced by the decisions of the user.  This ``personal equation''
includes a variety of factors that can vary from user to user, such as
how the continuum level is set, the line width in wavelength space,
and considerations for smoothing.  In order to obtain a consistent set
of measurements, all of these factors must be kept the same, usually
resulting in a single spectroscopist performing all measurements.  For
large scale, labor-intensive projects, such as CASH, it can be
difficult to remain consistent over hundreds of lines in hundreds of
stars.

This problem is somewhat resolved by the semi-automated methods, which
reduce the level of user interaction.  However, even in these
programs, the initial equivalent width can be flawed as the automated
continuum placement is rudimentary.  Weak lines and noise can be
indistinguishable, which can bias the continuum level calculated,
requiring user input into the continuum placement, which is generally
the largest component of error associated with the measurement.
Additionally, the line-finding algorithms often have difficulties in detecting
and measuring weak lines that are adjacent to large features.  

Fitline \citep{fitline1} is an example of a semi-automated equivalent
width measurement code.  It utilizes a genetic algorithm to fit
Gaussian profiles to the lines of a linelist by generating random
Gaussian parameters, mutating those parameters based on the lines that
are best fit, and iterating until convergence to the best fits.
Fitline requires some initial user interaction to dictate files.  It
can also be used with a post-processing code where the user can flag
fit quality. A detailed description of the code can be found in
\citet{fitline2}.

DAOSPEC \citep{daospec} is another equivalent width measurement
software that is written in FORTRAN, for use with IRAF libraries and
supermongo for graphical support. It works over five iterations to
determine the equivalent widths and in many ways is similar to ROBOSPECT,
using a line-fitting and continuum-fitting cycle.  DAOSPEC is not
interactive, although it does require the user to set input values,
including a FWHM estimate and the polynomial order used to fit the
continuum.

ROBOSPECT was written with the intent of being deterministic, therefore the
inputs required by ROBOSPECT do not greatly affect the resultant equivalent
width measurements.  It was also written to be user-friendly.  It
utilizes freely available libraries and graphics support so it is free
of cost as well as open source.  Additionally, since it is written in
C, it is not under any peculiar constraints in terms of the formatting
of input files and is easily modified for specific projects.

\section{Conclusions}\label{conclusions}

With the increasing scale of data in astronomical projects, automating
the process of measuring features is essential to make use of the
corresponding increase in computational power.  To support this goal
for the CASH project, we have written ROBOSPECT, a fully-automated
equivalent width measurement code. Although ROBOSPECT was designed for metal
poor stars, it is capable of fitting absorption and emission features
in a variety of astronomical sources.

Based on the results of a large number of simulated spectra with known
line parameters, we are confident that the fitting accuracy of ROBOSPECT is
consistent with the expected theoretical accuracy as presented by
\citet{eqw_err}.  These results are confirmed by the comparing the ROBOSPECT
results with manual measurements of both a sampling of CASH project
stars at a variety of S/N, as well as a set of long-slit spectra of
the Orion nebula.

While ROBOSPECT can be used to greatly reduce the labor involved in
determining the equivalent widths of spectral lines, it is not written
with the intention to be a substitute for actually looking at the
data.  Thus, a re-analysis of a subset of the lines that are visually
confirmed to be valid for stellar parameter determination and
abundance analysis shows an expected improvement in the scatter
relative to the manual measurements.  We found that our statistics
improved over all of our S/N values and for both of the ROBOSPECT parameter
settings we investigated.  In the case of the Hjerting model, we found
that extreme outliers, which would not be considered valid for stellar
abundance work, can bias the results.

Another caveat is that due to the method used to fit the lines and
measure the continuum, ROBOSPECT has difficulty with spectra in which there
is no discernible continuum over large wavelength regions.  In
metal-poor stars, these regions can be found in cool stars with
molecular bandheads, often from titanium and also in stars with
enhanced carbon abundances.  These features are also difficult to
define mathematically, preventing minimization process from finding a
single best fitting set of parameters.  Thus, ROBOSPECT should not be used
for these stars; however, most other routines will also have
difficulties with these stars.  Similarly, in the case of the most
carbon-enhanced stars, it is better to skip the equivalent width
measurement analysis in favor of a spectral synthesis.

\acknowledgements

JKH acknowledges support for this work from NSF grants AST 0708245 and
AST 0908978.  CZW acknowledges support for this work through the
PanSTARRS Science Consortium, which has been made possible through
contributions of the Institute for Astronomy, the University of
Hawaii, the Pan-STARRS Project Office, the Max-Planck Society and its
participating institutes, the Max Planck Institute for Astronomy,
Heidelberg and the Max Planck Institute for Extraterrestrial Physics,
Garching, The Johns Hopkins University, Durham University, the
University of Edinburgh, Queen's University Belfast, the
Harvard-Smithsonian Center for Astrophysics, and the Las Cumbres
Observatory Global Telescope Network, Incorporated, the National
Central University of Taiwan, and the National Aeronautics and Space
Administration under Grant No. NNX08AR22G issued through the Planetary
Science Division of the NASA Science Mission Directorate. Partial
support for this work was provided by National Science Foundation
grant AST-1009749.  We are very grateful to Jack Baldwin for supplying
the Orion nebula spectra used in this paper.  We also appreciate
useful conversations with Christopher Sneden, Harriet Dinerstein, Anna
Frebel, Matthew Shetrone, Randi Ludwig, and Eugene Magnier.

\clearpage

\begin{deluxetable}{lcc}
  \tablecolumns{3} 
  \tablewidth{0pc} 
  \tablecaption{CASH Comparison Stars}
  \tablehead{\colhead{Star}&\colhead{$S/N$}&\colhead{Expected Manual Equivalent Width Error}  \\
                \colhead{}&\colhead{}&\colhead{[m\AA{}]}}
\startdata
HD~122563 & 300 & 1.00 \\
HD~140283 & 100 & 3.00 \\
HD~115444 & 65 & 4.61 \\
HE~2137+0153 & 45 & 6.66 \\
HE~1225+0155 & 30 & 9.99 \\
\enddata 
\tablecomments{Information on the stars from \citet{cash2} used to test
  ROBOSPECT.}
\label{teststars}
\end{deluxetable} 

\begin{deluxetable}{ll}
  \tablecolumns{2} 
  \tablewidth{0pc} 
  \tablecaption{Default ROBOSPECT Settings Use for CASH Star Analysis}
  \tablehead{\colhead{Parameter}&\colhead{Description}}
\startdata
 continuum model & simple boxcar with a width of 40 m\AA{}\\
 line model & ``best'' estimating lines from FWHM and use deblending\\
 line profile function & Gaussian\\
 linelist & specify lines for measurement from linelist\\
 find lines & find additional lines to the fit outside of the linelist\\
\enddata 
\label{defaultparam}
\end{deluxetable} 

\begin{deluxetable}{lcccccccccc}
  \tablecolumns{11} 
  \tablewidth{0pc} 
  \tablecaption{Deviation Between Manual and ROBOSPECT Equivalent Width Measurements}
  \tablehead{\colhead{$S/N$}&\colhead{Iteration}&\colhead{Mean}&\colhead{$\sigma$}&\colhead{Median}&\colhead{Low}&\colhead{High}&\colhead{Q25}&\colhead{Q75}&\colhead{$\sigma_Q$}&\colhead{N}\\
\colhead{}&\colhead{}&\colhead{[m\AA{}]}&\colhead{[m\AA{}]}&\colhead{[m\AA{}]}&\colhead{[m\AA{}]}&\colhead{[m\AA{}]}&\colhead{[m\AA{}]}&\colhead{[m\AA{}]}&\colhead{[m\AA{}]}&\colhead{}}
  
\startdata
30 & 1 & 10.66 & 8.573 & 9.1 & -3.7 & 42.0 & 5.4 & 14.6 & 6.822 & 76 \\ 
30 & 2 & 9.411 & 8.458 & 8.0 & -5.5 & 40.0 & 4.1 & 11.7 & 5.635 & 79 \\ 
30 & 3 & 8.962 & 8.249 & 7.5 & -6.6 & 39.0 & 3.8 & 11.6 & 5.784 & 79 \\ 
30 & 4 & 8.799 & 8.264 & 7.5 & -6.8 & 38.7 & 3.6 & 11.6 & 5.932 & 79 \\ 
30 & 5 & 8.701 & 8.234 & 7.4 & -6.8 & 38.4 & 3.6 & 11.6 & 5.932 & 79 \\ 
45 & 1 & 5.544 & 5.198 & 5.5 & -6.2 & 22.6 & 2.0 &  7.6 & 4.152 & 55 \\ 
45 & 2 & 5.086 & 5.212 & 4.7 & -7.4 & 22.9 & 2.0 & 7.5 & 4.078 & 57 \\ 
45 & 3 & 5.141 & 5.110 & 4.4 & -5.3 & 22.9 & 2.1 & 7.5 & 4.004 & 58 \\ 
45 & 4 & 5.119 & 5.088 & 4.3 & -5.8 & 22.9 & 2.2 & 7.4 & 3.856 & 58 \\ 
45 & 5 & 5.095 & 5.024 & 4.4 & -6.0 & 22.8 & 2.3 & 7.4 & 3.782 & 59 \\ 
65 & 1 & 2.797 & 2.598 & 2.6 & -1.9 & 10.8 & 1.2 & 4 & 2.076 & 89 \\ 
65 & 2 & 2.051 & 2.573 & 1.7 & -3.1 & 10.7 & 0.4 & 3.6 & 2.373 & 90 \\ 
65 & 3 & 1.787 & 2.582 & 1.4 & -3.5 & 11.0 & 0.1 & 3.2 & 2.299 & 90 \\ 
65 & 4 & 1.700 & 2.572 & 1.3 & -3.7 & 11.4 & -0.1 & 3.2 & 2.447 & 90 \\ 
65 & 5 & 1.660 & 2.565 & 1.3 & -3.7 & 11.5 & -0.1 & 3.1 & 2.373 & 90 \\ 
100 & 1 & 1.437 & 2.457 & 1.2 & -7.4 & 7.2 & 0.0 & 2.9 & 2.150 & 80 \\ 
100 & 2 & 1.158 & 2.406 & 0.9 & -7.9 & 6.8 & -0.2 & 2.3 & 1.854 & 78 \\ 
100 & 3 & 1.111 & 2.435 & 0.8 & -8.0 & 6.6 & -0.3 & 2.5 & 2.076 & 81 \\ 
100 & 4 & 1.169 & 2.217 & 0.8 & -5.5 & 6.6 & -0.4 & 2.4 & 2.076 & 81 \\ 
100 & 5 & 1.157 & 2.213 & 0.8 & -5.4 & 6.5 & -0.4 & 2.3 & 2.002 & 81 \\ 
300 & 1 & 1.616 & 3.733 & 1.3 & -5.9 & 24.5 & 0.4 & 2.2 & 1.335 & 89 \\ 
300 & 2 & 1.048 & 3.215 & 0.6 & -5.7 & 20.4 & -0.2 & 1.4 & 1.186 & 90 \\ 
300 & 3 & 0.925 & 3.134 & 0.5 & -4.5 & 18.6 & -0.4 & 1.2 & 1.186 & 87 \\ 
300 & 4 & 0.802 & 3.009 & 0.4 & -5.5 & 17.6 & -0.5 & 1.3 & 1.335 & 89 \\ 
300 & 5 & 0.701 & 2.936 & 0.3 & -6.4 & 17.2 & -0.5 & 1.1 & 1.186 & 89 \\ 
\enddata
\tablecomments{Table containing the detailed statistics of the difference in
  equivalent widths measured between the manual measurements from
  \citet{cash2} and ROBOSPECT for the Gaussian line fits.  For each S/N, we
  show statistics for each iteration, including the mean, standard
  deviation ($\sigma$), the lowest value (low), highest value (high),
  25th Percentile value (Q25), 75th Percentile value (Q75), the $\sigma_Q$
  value, and the number of lines from which these statistics were
  derived.}
\label{gaussanal}
\end{deluxetable} 

\begin{deluxetable}{lcccccccccc}
  \tablecolumns{11} 
  \tablewidth{0pc} 
  \tablecaption{Deviation Between Manual and ROBOSPECT Equivalent Width Measurements: Hjerting Model}
  \tablehead{\colhead{$S/N$}&\colhead{Iteration}&\colhead{Mean}&\colhead{$\sigma$}&\colhead{Median}&\colhead{Low}&\colhead{High}&\colhead{Q25}&\colhead{Q75}&\colhead{$\sigma_Q$}&\colhead{N}\\
\colhead{}&\colhead{}&\colhead{[m\AA{}]}&\colhead{[m\AA{}]}&\colhead{[m\AA{}]}&\colhead{[m\AA{}]}&\colhead{[m\AA{}]}&\colhead{[m\AA{}]}&\colhead{[m\AA{}]}&\colhead{[m\AA{}]}&\colhead{}}
  
\startdata
30 & 1 & 10.51 & 8.923 & 9.4 & -14.5 & 42.4 & 5.8 & 13.1 & 5.413 & 76 \\ 
30 & 2 & 9.647 & 8.864 & 8.5 & -16.1 & 40.4 & 4.6 & 13.5 & 6.599 & 79 \\ 
30 & 3 & 9.756 & 9.616 & 8.4 & -16.7 & 45.3 & 4.4 & 13.3 & 6.599 & 79 \\ 
30 & 4 & 9.222 & 8.167 & 7.7 & -5.9 & 39.6 & 3.6 & 11.7 & 6.006 & 79 \\ 
30 & 5 & 9.08 & 8.135 & 7.8 & -6.5 & 39 & 4 & 11.7 & 5.71 & 79 \\ 
45 & 1 & 5.236 & 6.029 & 5.3 & -16.2 & 22.7 & 1.9 & 8.1 & 4.597 & 56 \\ 
45 & 2 & 5.095 & 5.46 & 4.8 & -7.7 & 22.8 & 2.1 & 7.5 & 4.004 & 57 \\ 
45 & 3 & 5.128 & 5.604 & 5.2 & -10.4 & 22.9 & 2.5 & 7.7 & 3.856 & 57 \\ 
45 & 4 & 5.65 & 5.849 & 5.2 & -8.9 & 25.7 & 2.6 & 7.7 & 3.782 & 58 \\ 
45 & 5 & 5.268 & 5.826 & 5.1 & -10.3 & 22.8 & 2.5 & 7.7 & 3.856 & 59 \\ 
65 & 1 & 2.862 & 3.164 & 2.8 & -13.1 & 11.6 & 1.3 & 4.3 & 2.224 & 89 \\ 
65 & 2 & 2.309 & 4.496 & 1.8 & -8.8 & 35.1 & 0.3 & 3.7 & 2.521 & 87 \\ 
65 & 3 & 1.8 & 2.502 & 1.5 & -3.4 & 9 & 0.1 & 3.7 & 2.669 & 84 \\ 
65 & 4 & 1.89 & 2.51 & 1.6 & -3.6 & 9.1 & 0.1 & 3.7 & 2.669 & 89 \\ 
65 & 5 & 1.32 & 4.242 & 1.4 & -30.5 & 9 & 0 & 3.5 & 2.595 & 88 \\ 
100 & 1 & 1.575 & 2.595 & 1.2 & -8.1 & 8 & 0.1 & 3.1 & 2.224 & 80 \\ 
100 & 2 & 1.204 & 2.657 & 1 & -8.4 & 8.5 & -0.2 & 2.5 & 2.002 & 80 \\ 
100 & 3 & 1.146 & 2.526 & 1 & -8.1 & 7.6 & -0.2 & 2.6 & 2.076 & 81 \\ 
100 & 4 & 1.068 & 2.524 & 0.9 & -8.5 & 6.6 & -0.2 & 2.4 & 1.928 & 82 \\ 
100 & 5 & 0.989 & 2.618 & 0.8 & -8.5 & 6.6 & -0.4 & 2.4 & 2.076 & 82 \\ 
300 & 1 & -0.004 & 8.456 & 1.3 & -47.9 & 24.6 & 0.4 & 2.1 & 1.261 & 90 \\ 
300 & 2 & 1.147 & 5.64 & 0.6 & -19.3 & 20.6 & -0.1 & 1.6 & 1.261 & 90 \\ 
300 & 3 & 2.414 & 8.438 & 0.6 & -19.6 & 55.7 & -0.3 & 2 & 1.705 & 85 \\ 
300 & 4 & 2.533 & 11.03 & 0.6 & -33.2 & 56.5 & -0.4 & 1.8 & 1.631 & 88 \\ 
300 & 5 & 1.912 & 5.247 & 0.5 & -9.7 & 26.6 & -0.3 & 1.9 & 1.631 & 89 \\ 
\enddata 
\tablecomments{Table containing the detailed statistics of the difference in
  equivalent widths measured between the manual measurements from
  \citet{cash2} and ROBOSPECT for the Hjerting line profile.  The columns
  are the same as in Table~\ref{gaussanal}}
\label{hjertinganal}
\end{deluxetable} 

\begin{deluxetable}{lcccccccccc}
  \tablecolumns{11} 
  \tablewidth{0pc} 
  \tablecaption{Equivalent Width Deviations Using A Restricted Set of Lines}
  \tablehead{\colhead{$S/N$}&\colhead{Setting}&\colhead{Mean}&\colhead{$\sigma$}&\colhead{Median}&\colhead{Low}&\colhead{High}&\colhead{Q25}&\colhead{Q75}&\colhead{$\sigma_Q$}&\colhead{N}\\
\colhead{}&\colhead{}&\colhead{[m\AA{}]}&\colhead{[m\AA{}]}&\colhead{[m\AA{}]}&\colhead{[m\AA{}]}&\colhead{[m\AA{}]}&\colhead{[m\AA{}]}&\colhead{[m\AA{}]}&\colhead{[m\AA{}]}&\colhead{}}
\startdata
300 & Gaussian &  0.445 & 2.240 & 0.2 &  -6.4 &  8.4 & -0.5 & 1.0 & 1.112 &  82 \\ 
300 & Hjerting &  0.654 & 1.900 & 0.3 &  -3.6 &  8.3 & -0.3 & 1.2 & 1.112 &  74 \\ 
100 & Gaussian &  0.881 & 2.112 & 0.5 &  -5.4 &  6.5 & -0.5 & 1.8 & 1.705 &  72 \\ 
100 & Hjerting &  0.496 & 2.220 & 0.4 &  -8.5 &  6.4 & -0.6 & 1.4 & 1.483 &  68 \\ 
65  & Gaussian &  1.421 & 2.273 & 1.2 &  -3.7 &  8.7 & -0.2 & 2.7 & 2.150 &  85 \\ 
65  & Hjerting &  1.353 & 2.016 & 1.1 &  -2.9 &  6.0 &  0.0 & 2.6 & 1.928 &  76 \\ 
45  & Gaussian &  4.747 & 4.828 & 4.2 &  -6.0 & 22.8 &  2.1 & 7.4 & 3.930 &  55 \\ 
45  & Hjerting &  3.808 & 4.817 & 4.1 & -10.3 & 13.1 &  2.0 & 6.8 & 3.559 &  48 \\ 
30  & Gaussian &  6.854 & 5.716 & 6.7 &  -5.3 & 25.7 &  3.1 & 9.8 & 4.968 &  68 \\ 
30  & Hjerting &  6.602 & 5.859 & 6.2 &  -6.5 & 26.1 &  3.0 & 9.6 & 4.894 &  63 \\ 
\enddata
\caption{Table containing the detailed statistics of the difference in
  equivalent widths measured between the manual measurements from
  \citet{cash2} and ROBOSPECT for line fits including only those lines which
  would be considered fit for an abundance analysis.  For each S/N, we
  show the same statistics as in Table~\ref{gaussanal} for the final
  iteration using both of the fitting parameters
  (Gaussian and Hjerting line profiles).}
\label{cookedanal}
\end{deluxetable} 

\begin{deluxetable}{lccccc}
  \tablecolumns{5} 
  \tablewidth{0pc} 
  \tablecaption{Orion spectra grating data}
  \tablehead{\colhead{Grating}&
    \colhead{Blaze}&
    \colhead{Wavelength}&
    \colhead{Resolution}&
    \colhead{Median Relative}&
    \colhead{Relative Flux}
      \\
      &
    \colhead{Wavelength (\AA{})}&
    \colhead{Coverage (\AA{})}&
    \colhead{FWHM (\AA{})}&
    \colhead{Flux Difference (\%)}&
    \colhead{Difference Scatter (\%)
    }    
  }
\startdata
Blue & 5000 & 6900 & 17.2 & 0.36 & 2.48 \\
High & 6750 & 1725 & 4.3  & -1.32 & 2.20 \\
Red  & 8000 & 6550 & 16.4 & -1.03 & 3.26 \\
\enddata 

\label{tab:orion1.5m}
\end{deluxetable}

\begin{thebibliography}{}
\bibitem[Cayrel(1988)]{eqw_err} Cayrel, R.\ 1988, The Impact of 
Very High S/N Spectroscopy on Stellar Physics, 132, 345 
\bibitem[Cayrel et al.(2004)]{cayrel2004} Cayrel, R., Depagne, E., Spite, M., et al.\ 2004, \aap, 416, 1117 
\bibitem[Fran{\c c}ois et al.(2003)]{fitline1} Fran{\c c}ois, P., Depagne, E., Hill, V., et al.\ 2003, \aap, 403, 1105 
\bibitem[Fulbright(2000)]{fulbright00} Fulbright, J.~P.\ 2000, \aj, 120, 1841 
\bibitem[Harris(1948)]{hjerting_reference} Harris, D.~L., III 1948, \apj, 108, 112
\bibitem[Hollek et al.(2011)]{cash2} Hollek, J.~K., Frebel, 
  A., Roederer, I.~U., et al.\ 2011, \apj, 742, 54 
\bibitem[Lee et al.(2008)]{sspp} Lee, Y.~S., Beers, T.~C., Sivarani, T., et al.\ 2008, \aj, 136, 2022 
\bibitem[Lemasle et al.(2007)]{fitline2} Lemasle, B., Fran{\c c}ois, P., Bono, G., et al.\ 2007, \aap, 467, 283 
\bibitem[Ludwig et al.(2012)]{agn} Ludwig, R.~R., Greene, 
J.~E., Barth, A.~J., \& Ho, L.~C.\ 2012, \apj, 756, 51 
\bibitem[Pellegrini et al.(2010)]{h2r} Pellegrini, E.~W., 
Baldwin, J.~A., \& Ferland, G.~J.\ 2010, \apjs, 191, 160 
\bibitem[Roederer \& Lawler(2012)]{ew} Roederer, I.~U., \& Lawler, J.~E.\ 2012, \apj, 750, 76 
\bibitem[Shetrone et al.(2007)]{het} Shetrone, M., Cornell, 
M.~E., Fowler, J.~R., et al.\ 2007, \pasp, 119, 556
\bibitem[Sterling et al.(2005)]{pne} Sterling, N.~C., 
Dinerstein, H.~L., Bowers, C.~W., \& Redfield, S.\ 2005, \apj, 625, 368 
\bibitem[Stetson \& Pancino(2008)]{daospec} Stetson, P.~B., \& Pancino, E.\ 2008, \pasp, 120, 1332 
\bibitem[Tull(1998)]{hrs} Tull, R.~G.\ 1998, \procspie, 
3355, 387 
\bibitem[Wallace et al.(2011)]{wallace_solar} Wallace, L., Hinkle, 
K.~H., Livingston, W.~C., \& Davis, S.~P.\ 2011, \apjs, 195, 6 
\bibitem[Yanny et al.(2009)]{segue} Yanny, B., Rockosi, C., Newberg, H.~J., et al.\ 2009, \aj, 137, 4377 
\bibitem[York et al.(2000)]{sdss} York, D.~G., Adelman, J., Anderson, J.~E., Jr., et al.\ 2000, \aj, 120, 1579 
\end{thebibliography}
\end{document}